 \documentclass[final,3p,times]{elsarticle}
\usepackage{babel,tabularx,ragged2e,booktabs}
\usepackage{lineno}

\usepackage[utf8]{inputenc}
\usepackage{textcomp}
\usepackage{graphicx}
\usepackage{amsmath}
\usepackage{siunitx}
\usepackage{longtable,tabularx}
\usepackage{color}
\usepackage{cancel}
\usepackage{color}
\usepackage{tcolorbox}
\usepackage{float}
\usepackage{multirow}
\usepackage{caption}
\usepackage{subcaption}
\usepackage{graphicx}

\usepackage{bm}
\usepackage[colorlinks]{hyperref}

\journal{}

\begin{document}

\begin{frontmatter}



\title{Advances in Fine Line-Of-Sight Control for Large Space Flexible Structures}

\author[ISAE]{Francesco Sanfedino\fnref{label2}}
\ead{francesco.sanfedino@isae-supaero.fr}
\fntext[cor1]{Corresponding author; Associate Professor}

\author[ISAE]{Gabriel Thiébaud}
\ead{gabriel.thiebaud@student.isae-supaero.fr}

\author[ISAE]{Daniel Alazard\fnref{label3}}
\ead{daniel.alazard@isae-supaero.fr}
\fntext[label3]{Professor}

\author[TAS]{Nicola Guercio}
\ead{nicola.guercio@thalesaleniaspace.com}

\author[ESA]{Nicolas Deslaef}
\ead{nicolas.deslaef@esa.int}

\address[ISAE]{Institut Supérieur de l’Aéronautique et de l’Espace (ISAE-SUPAERO), Université de Toulouse, 10 Avenue Edouard Belin, BP-54032, 31055, Toulouse
Cedex 4, France}
\address[TAS]{Thales Alenia Space, Cannes, Provence-Alpes-Côte d'Azur, 06150, France}
\address[ESA]{ESA-ESTEC, Noordwijk, South Holland, 2201 AZ, Netherlands}

\begin{abstract}

	The increased need in pointing performance for Earth observation and science Space missions together with the use of lighter and flexible structures directly come with the need of a robust pointing performance budget from the very beginning of the mission design. An extensive understanding of the system physics and its uncertainties is then necessary in order to push control design to the limits of performance and constrains the choice of the set of sensors and actuators.
A multi-body framework, the Two Input Two Output Ports approach, is used to build all the elementary flexible bodies and mechanisms involved in a fine pointing mission.  
This framework allows the authors to easily include all system dynamics with an analytical dependency on varying and uncertain mechanical parameters in a unique Linear Fractional Transformation (LFT) model. This approach opens the doors to modern robust control techniques that robustly guarantee the expected fine pointing requirements.
In particular, a novel control architecture is proposed to reduce the microvibrations induced both by reaction wheel imbalances and Solar Array Drive Mechanism driving signal, by letting them work during the imaging phase. Thanks to a set of accelerometers placed at the isolated base of the payload and in correspondence  of the mirrors with the largest size in a Space telescope (typically the primary and secondary ones), it is possible to estimate the line-of-sight error at the payload level by hybridizing them with the low-frequency measurements of the camera. While a classical Fast Steering Mirror in front of the camera can compensate  for a large amount of microvibration, an innovative architecture with a set of six Proof-Mass Actuators installed at the payload isolator level can further improve the pointing performance.
In particular, it is shown how the proposed architecture is able to robustly guarantee an absolute performance error of 10 arcsec in face of system parametric uncertainties at low frequency ($\approx$ 1 rad/s) with a progressive reduction of the jitter down to 40 marcsec for higher frequencies where micro-vibration sources act.

\end{abstract}


\end{frontmatter}

%\linenumbers


\section*{Nomenclature}

{
\noindent\begin{longtable*}{@{}l @{\quad \quad} l@{}>{\(\collectcell\mathbf}l<{\endcollectcell\)}}
$\mathcal{R}_b$ & Main body $\mathcal{B}$ reference frame: $\mathcal{R}_b = (O,\mathbf{x}_b,\mathbf{y}_b,\mathbf{z}_b)$. $O$ is a reference point on the main body.\\
$\mathcal{R}_a$ &   Appendage $\mathcal{A}$ reference frame: $\mathcal{R}_a = (P,\mathbf{x}_a,\mathbf{y}_a,\mathbf{z}_a)$. $P$ is the appendage's anchor point on the\\
 & main body.\\
$\mathbf{P}_{a/b}$ & Direction cosine matrix of the rotation from frame $\mathcal{R}_b$ to frame $\mathcal{R}_a$.\\
$\mathbf{a}_P$ & Linear acceleration (vector) of body $\mathcal{B}$ at point $P$, expressed in \si{\meter\per\square\second}.\\
$\boldsymbol{\omega}$ & Angular velocity (vector) of $\mathcal{R}_b$ with respect to the inertial frame, expressed in \si{\radian\per\second}.\\
$\bm{\ddot{\mathbf{x}}}_P$ & acceleration twist at point $P$: $\bm{\ddot{\mathbf{x}}}_P=[\mathbf{a}_P^T\;\;\bm{\dot{\omega}}^T]^T$, expressed in $\left[\si{\meter\per\square\second}\;\; \si{\radian\per\square\second}\right]$.\\
$\mathbf{F}_{ext}$  & External forces (vector) applied to  body $\mathcal{B}$ expressed in \si{\newton}.\\
$\mathbf{T}_{ext,B}$ & External torques (vector) applied to body $\mathcal{B}$ at point $B$, expressed in \si{\newton\meter}.\\
$\mathbf{F}_{\mathcal{B/A}}$ & Force (vector) applied by body $\mathcal{B}$ on body $\mathcal{A}$, expressed in \si{\newton}.\\
$\mathbf{T}_{\mathcal{B/A},P}$ & Torque (vector) applied by body $\mathcal{B}$ on body $\mathcal{A}$ at point $P$, expressed in \si{\newton\meter}.\\
$\mathbf{W}_{\mathcal{B/A},P}$ & Wrench applied by body $\mathcal{B}$ on body $\mathcal{A}$ at point $P$: $\mathbf{W}_{\mathcal{B/A},P}=[\mathbf{F}^T_{\mathcal{B/A}}\;\;\mathbf{T}^T_{\mathcal{B/A},P}]^T$ expressed\\
& in $\left[\si{\newton}\;\; \si{\newton\meter}\right]$.\\
$\mathcal{M}^{\mathcal{A}}_{PC}(\mathrm{s})$ & TITOP model of appendage $\mathcal{A}$ at parent point $P$ and child point $C$.\\
$\mathcal{F}_l(\mathbf{G}\left(\mathrm{s}\right),\mathbf{K}(\mathrm s))$ & Lower LFT of the system $\mathbf{G}\left(\mathrm{s}\right)$ and the controller $\mathbf{K}(\mathrm s)$.\\
$\mathcal{F}_u(\mathbf{G}\left(\mathrm{s}\right),\bm{\Delta})$ & Upper LFT of the system $\mathbf{G}\left(\mathrm{s}\right)$ and the uncertain block $\bm{\Delta}$.\\
$\boldsymbol{\tau}_{PB}$ &  Kinematic model between points $P$ and $B$: %\hspace{3cm} 
$\boldsymbol{\tau}_{PB}=\left[\begin{array}{cc}
	\mathbf{I}_3  & (^*\overrightarrow{PB})\\
	\mathbf{0}_{3\times 3} & \mathbf{1}_3  \\
\end{array}\right]$.\\
$(^*\overrightarrow{PB})$ & Skew symmetric matrix associated with vector $\overrightarrow{PB}$.\\
$\left[\mathbf{X}\right]_{\mathcal{R}_i}$ & $\mathbf{X}$ (model, vector or tensor) expressed in frame $\mathcal{R}_i$.\\
$\mathbf{v}\{i\}$ & Component $i$ of vector $\mathbf{v}$\\
$\frac{d \mathbf{v}}{dt}|_{\mathcal{R}_i}=0$ & Derivative of vector $\mathbf{v}$ with respect to frame $\mathcal{R}_i$.\\
$\mathbf{u} \times \mathbf{v}$& Cross product of vector $\mathbf{u}$ with vector $\mathbf{v}$ $\left(\mathbf{u} \times \mathbf{v}=(^*\mathbf{u})\mathbf{v}\right)$\\
$\mathbf{u}.\mathbf{v}$ & Dot product of vector $\mathbf{u}$ with vector $\mathbf{v}$ ($\mathbf{u}.\mathbf{v}=\left[\mathbf{u}\right]_{\mathcal{R}_i}^T\left[\mathbf{v}\right]_{\mathcal{R}_i},\;\forall\,\mathcal{R}_i$)\\
$\mathrm{s}$ &  \textsc{Laplace}'s variable.\\
$\mathbf{I}_n$ &  Identity matrix $n \times n$.\\
$\mathbf{0}_{n\times m}$  & Zero matrix $n \times m$.\\
$\mathbf{A}\{i,j\}$ & Element $(i,j)$ of matrix $\mathbf{A}$\\
$\mathbf{A}^\mathrm T$ & Transpose of $\mathbf{A}$.\\
diag$(\omega_i)$ & Diagonal matrix $N\times N$: $\mbox{diag}(\omega_i)\{i,i\}=\omega_i$, $i=1,\cdots, N$.\\
$\operatorname{null}\left(\mathbf{A}\right)$ & Null space of the matrix $\mathbf{A}$.\\
$\operatorname{blkdiag}\left(\mathbf{A}_1,\mathbf{A}_2\right)$ & Block-diagonal assembly of $\mathbf{A}_1$ and $\mathbf{A}_2$.
\end{longtable*}}

\section{Introduction}
With the development of the next generation of Earth observation and science Space missions, there is an increasing trend towards highly performing payloads. This trend is leading to increased detector resolution and sensitivity, as well as longer integration time which directly drive pointing requirements to higher stability and lower line-of-sight (LOS) jitter \cite{nasa_microvib}. Such instruments typically comes with stringent pointing requirements and constraints on attitude and rate stability over an extended frequency range well beyond the attitude control system (ACS) bandwidth, by entailing micro-vibration mitigation down to the arcsecond (arcsec) level or less \cite{large_st_iso}\cite{euclid_design}. Micro-vibrations are defined in \cite{ECSS_E_HB_32_26A} as low-level vibrations causing a distortion of the LOS during on-orbit operations of mobile or vibratory parts. Two main classes of spacecraft disturbance sources have been indeed identified in \cite{ECSS_E_HB_32_26A}: external or natural events (micro-meteorids and debris impacts, atmospheric drag, Earth gravity field gradient, Earth magnetic field, solar flux and Earth albedo, eclipse entry and exit) and internal events (propulsion subsystem, avionics subsystem, electrical power subsystem, radio frequency/telemetry and telecommand subsystem, thermal control subsystem, structure subsystem). Disturbances induced by internal elements and propagating along the spacecraft large flexible appendages are the main contributor to micro-vibration pointing budget. Among this class of disturbances, two types of internal disturbances can be distinguished  from their frequency content: \textit{periodic} or \textit{harmonic} disturbances (occurring for long-time duration with a spectrum generally characterized by a fundamental frequency component and several other harmonics that are integer or non-integer multiple of it) and \textit{transient} disturbances. Harmonic disturbances can be caused by: reaction wheels (RW) \cite{doi:10.2514/6.2007-6736}, control momentum gyros (CMG), gyroscopes, solar array drive mechanisms (SADM) \cite{10.1117/12.18234, ZHANG2021106398} if continuously actuated, cryogenic coolers \cite{doi:10.1063/1.3553198}, heat pipes. On the other hand in the category of transient disturbances it can be found: SADM when actuated in particular mission phases, antenna pointing mechanism (APM) \cite{640772}, mirror scanning mechanisms \cite{SUDEY1985485}, micro-thrusters, gas flow regulators, latch valves, relays, sudden stress release, clank phenomena. Reaction wheel static and dynamic imbalances, bearing and motor imperfections, and stepper motors within SADM are among the most important sources of periodic micro-vibrations. Many studies have focused on the characterization and the development of empirical and analytic models of the SADM \cite{SANFEDINO2022108168} and RW disturbances \cite{doi:10.2514/6.1999-1204}, \cite{MASTERSON2002575}, \cite{KIM20144214} coupled with the natural modes of the spacecraft main structure \cite{ZHANG20135748,ADDARI2017225,sanfe2019,preda:tel-01722860,preda:hal-01887323}. Micro-vibrations are in fact propagated to the sensitive payload along the spacecraft structure and amplified by its flexibility. A combination of passive and active control techniques is used in \cite{preda:tel-01722860} to achieve wide-band micro-vibration mitigation: passive isolation is employed to reduce high-frequency modes and noise, while active mechanisms such as proof-mass actuators (PMA) actuated in closed-loop are used to control the low to medium frequency range (typically from a few Hz up to a few hundred Hz) above the ACS bandwidth. This hybrid isolation system can be integrated either at the disturbance source level (as done in \cite{preda:hal-01887323} where both passive and active control solutions are applied at the base of the RW) or directly at the payload level. Yun \textit{et al.} \cite{YUN2020105543} combined a Stewart platform and a piezoelectric Fast Steering Mirror (FSM) in a two-stage stabilization platform of the optical payload. Zhang $\textit{et al.}$ \cite{ZHANG201899} proposed an active isolated platform based on magnetic suspensions.

However, in order to guarantee high pointing performance, it is necessary to entirely characterize the transmission path between the micro-vibration source and the payload. The earlier the model is available, the easier it is to meet the stringent pointing requirements, by designing appropriate control strategies. The main difficulties encountered in Space system characterization are both the impossibility to correctly identify the system on ground due to the presence of gravity and the consideration of all possible system uncertainties. Several uncertainties are in fact determined by: manufacture imperfections of structures and mechanisms, evolution of the system during the mission (i.e. material exposition to Space environment, mass and inertia variation due to ejected ergol), misknowledge of sensors/actuator dynamics. Uncertainty quantification is the preliminary step to be accomplished before designing robust control laws which provide a certificate on the closed-loop system stability and performance. NASA Langley Research Center \cite{crespo2010computational,crespo2014nasa} has recently contributed to this topic by highlighting the role played by system uncertainty on control performance.

Once the plant's dynamics is modeled together with its uncertainties, modern robust control techniques can then be used to synthesize fixed-structured controllers that meet the pointing requirements for a large set of system uncertainties while maintaining low complexity for practical implementation \cite{apk2014,Apkarian2015}.

The impact of control/structure interactions has been highlighted by several research studies in the 90’s. Belvin \textit{et al.} \cite{Belv95} considered the application of passive and active payload mounts for attenuation of pointing jitter of the EOS AM-1 spacecraft. O'Brien \textit{et al.} \cite{Obrien95} showed the problem of the isolation of a spaceborne interferometer.
Miller \textit{et al.}  \cite{931545} modeled the impact of the broadband and narrowband disturbance of the flywheel and bearing imperfections on the pointing performance of the Space Interferometry Mission (SIM).
The widespread technique to numerically model complex industrial flexible dynamic systems is the Finite Element Method (FEM). However, while it is necessary to finely characterize the system behavior with all its uncertainties, the complexity of the plant (i.e. number of states and uncertainty occurrences) has to be kept small in order to make this plant exploitable for control synthesis and practical implementation. 
FEM models strictly used for structure assessment \cite{aglietti2019}, characterized by thousands of degrees-of-freedom (DOFs), cannot be directly exploited for control synthesis and analysis and need to be properly reduced \cite{931545}. Moreover, a classic nominal FEM model-based controller \cite{GASBARRI2014252} suffer from a lack of uncertainty characterization and its validation implies time consuming Monte Carlo's simulations, that can skip rare but critical worst-case configurations.

The possibility to take into account parametric variations in a model fully compatible with the standard robust analysis and control tools opens new insights to design and prototype spacecraft architectures while taking into account all the subsystems (structure modelling, control, optics, mechanism disturbances).

In this spirit the Two-Input Two-Output Ports (TITOP) approach, firstly proposed by Alazard \textit{et al.} \cite{titop} and further extended by Chebbi \textit{et al.} \cite{lin_dyn_flex} and Sanfedino \textit{et al.} \cite{SANFEDINO2018128}, offers the opportunity to assemble several flexible sub-structures by keeping the analytical dependency of the overall model on the constitutive mechanical parameters and reducing this dependency to the minimal number of occurrences. This multi-body approach has been conceived in order to perfectly fit with the Linear Fractional Transformation (LFT) theory developed in the robust control framework \cite{zhou1998essentials}. It is in fact possible to include any kind of uncertain and varying parameters with a minimum number of occurrences and recover the dynamic (forces and torques) and kinematic (linear and angular accelerations, speeds, displacements) quantities at the connection nodes of each body. In this way, a huge family of possible plants can be incorporated in a unique LFT model that informs the control synthesis algorithm of all possible uncertain and varying parameters.
All substructure models derived for simple (i.e. beams and plates) or complex (FEM models of 3D industrial bodies) geometries and mechanisms have been integrated in a MATLAB/Simulink environment using the Satellite Dynamics Toolbox (SDT), a collection of ready-to-use blocks that allows rapid prototyping of complex multi-body systems for space applications \cite{sdt}\cite{sdt_user_guide}. The resulting spacecraft model is then ready for robust control synthesis and robust stability and performance assessment by using the MATLAB routines available in the Robust Control Toolbox \cite{RCT_user_guide}. Thanks to the TITOP formalism Perez \textit{et al.} \cite{perez2015linear} showed how to perform a simultaneous ACS/structure co-design of a large flexible spacecraft, Sanfedino \textit{et al.} \cite{SANFEDINO2022108168} designed a robust estimation filter for on-orbit micro-vibrations characterization of a SADM, Finozzi \textit{et al.} \cite{FINOZZI2022109427} designed an optimal truss-structure, actively controlled by PMAs for a high accuracy pointing antenna.

The first contribution of this paper is to present a set of novel dynamic models of mechanisms (namely RW, FSM, PMA) in the TITOP framework, which are typically involved in a modern science mission design (i.e. a Space telescope). An industrial benchmark is then assembled for LOS robust control design.

The second contribution of this work is to outline an innovative control architecture that combines multiple passive/active control strategies in order to get a considerable mitigation of the microvibrations induced by RW imbalances and SADM input signals. In particular, combining the measurements of accelerometers in key points of the structure (at the payload isolator base and at mirrors $M_1$ and $M_2$ of the space telescope) and the low frequency acquisitions of a camera, the LOS is directly corrected by an FSM and indirectly by a set of PMAs installed at the payload isolator level. This strategy allows to get fine pointing performance while leaving the RW and SADM still working during the imaging phase with consequent increase of the time window available for the scientific observations. Indeed, one of the classical but constraining strategies for this kind of missions is to interrupt the operation of the micro-vibration sources in order to improve the pointing performance while having an impact on their primary functionalities (i.e. attitude control for RW, optimization of the received Sun power for the SADM).

The final goal of this paper is to show how to finely model an industrial flexible spacecraft with tight pointing requirements and design a robust controller able to push the potential achievable performance to its limits by coping with modeled system uncertainties.

This paper is organized in three main parts: modeling of a fine LOS platform with all micro-vibration sources and active control actuators, robust control architecture for fine pointing performance, analysis of the final system performance.
Section \ref{titop_sec} presents the main elements of the TITOP theory. The modeling of the micro-vibration sources and actuators is then tackled in section \ref{mv_sources_sec} and \ref{mv_act_sec}. The full spacecraft dynamic model is outlined in section \ref{sc_dyn_sec} by assembly of sub-systems using the TITOP framework. Section \ref{fine_los_control} presents the two novel robust control architectures proposed in this work, one relaying on the estimation of the LOS and compensation with an FSM; and the second one based on a set of PMAs controlling an active/passive isolator system placed at the base of the payload. In section \ref{results_discussion}, results of the analysis are showcased and discussed. Finally, some conclusions and remarks are presented in section \ref{conclusion}.

\section{Modeling}

\subsection{The TITOP framework}
\label{titop_sec}

\begin{figure}[h!]
	\centering
	\includegraphics[width=\columnwidth]{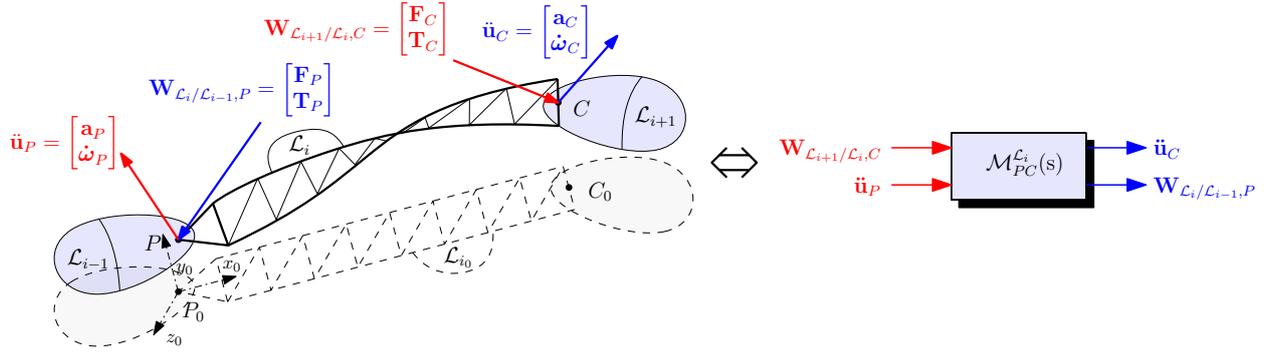}
	\caption{i-th flexible appendage sub-structured body (left) and equivalent TITOP model (right)}
	\label{fig:titop_beam}
\end{figure}

Let us consider the generic flexible appendage $\mathcal{L}_i$ in Fig. \ref{fig:titop_beam} linked to a parent substructure $\mathcal{L}_{i-1}$ at point $P$ and to a child substructure $\mathcal{L}_{i+1}$ at point $C$. 
Moreover let us define the reference frame $\mathcal{R}_0 = (P, x_0, y_0, z_0)$ centered in node P of $\mathcal{L}_i$ in equilibrium conditions.
In the model of the appendage $\mathcal{L}_i$, clamped-free boundary conditions are considered: the joint at point P is rigid and statically determinate, with the parent body
$\mathcal{L}_{i-1}$ imposing a motion on $\mathcal{L}_i$, while point $C$ is internal and unconstrained, and the action of $\mathcal{L}_{i+1}$ on $\mathcal{L}_{i}$ is a contact effort.

 The TITOP model $\mathcal{M}_{PC}^{\mathcal{L}_i}(\mathrm{s})$ is a linear state-space model with 12
inputs (6 for each of the two input ports):

\begin{enumerate}
	\item The 6 components in $\mathcal{R}_0$ of the wrench $\mathbf{W}_{\mathcal{L}_{i+1}/\mathcal{L}_i,C}$ composed of the three-components
	force vector $\mathbf{F}_C$ and the three-components torque vector $\mathbf{T}_C$ applied by $\mathcal{L}_{i+1}$ to $\mathcal{L}_i$ at
	the free node $C$;
	\item The 6 components in $\mathcal{R}_0$ of the acceleration vector $\ddot{\mathbf{u}}_P$ composed of the three-components
	linear acceleration vector $\mathbf{a}_P$ and the three-components angular acceleration vector $\dot{\bm{\omega}}_P$
	at the clamped node $P$;
\end{enumerate}

\noindent and 12 outputs (6 for each of the two output ports):

\begin{enumerate}
	\item The 6 components in $\mathcal{R}_0$ of the acceleration vector $\ddot{\mathbf{u}}_C$ at the free node $C$;
	\item The 6 components in $\mathcal{R}_0$ of the wrench $\mathbf{W}_{\mathcal{L}_i/\mathcal{L}_{i-1},P}$ applied by $\mathcal{L}_i$ to the parent structure
	$\mathcal{L}_{i-1}$ at the clamped node $P$.
\end{enumerate}

The TITOP model $\mathcal M^{\mathcal L_i}_{P,C}(\mathrm s)$ displayed in Fig. 1 (right) includes in a minimal state-space model the direct dynamic model (transfer from acceleration twist to wrench) at point $P$ and the inverse dynamic model (transfer from wrench to acceleration twist) at point $C$.

This model, conceived with the clamped-free condition, is useful to study any other kind of boundary configuration as proven in \cite{lin_dyn_flex} thanks to the invertibility of all of its 12 input-output channels.

\subsection{Modeling of micro-vibration sources}
\label{mv_sources_sec}

Among all possible sources of micro-vibration, internal events and more particularly disturbances from rotating mechanisms (like reaction wheels and control momentum gyros) and electrical drive units of gimballed mechanisms (like SADM, APM and mirror scanning mechanisms), are the biggest contributors to the pointing error budget \cite{sanfe2019}. The high-speed rotation of reaction wheels produces harmonic force and torque disturbances on the parent structure. It has been identified that those perturbations are caused by the rotating static and dynamic mass imbalances of the wheel, the ball bearing imperfections, as well as the motor imperfections (torque ripple and motor cogging) \cite{billbialke}, all functions of the flywheel angular velocity. On the other hand, gimbaled mechanisms are typically driven by stepper motors that are actuated by harmonic voltage signals. Such signals get transmitted by the electro-mechanical conversion and propagated to the structure through the mechanism. These two main micro-vibration sources are considered in this study:

\begin{enumerate}
	\item The reaction wheel system (RWS) disturbances;
	\item The solar array drive mechanism (SADM) disturbances;
\end{enumerate}

The solar array (SA) and SADM models are detailed in \cite{SANFEDINO2022108168}. The RWS is modeled hereafter in the TITOP framework.

\begin{figure}[h!]
	\centering
	\includegraphics[scale=0.6]{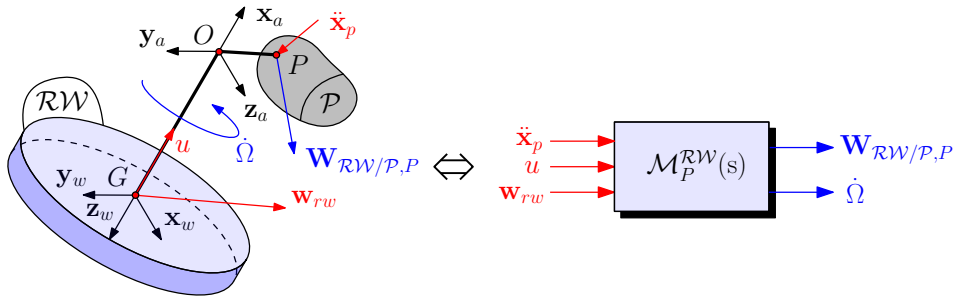}
	\caption{Single reaction wheel (left) and equivalent TITOP model (right)}
	\label{single_rw}
\end{figure}

Let us consider the single spinning wheel depicted in Fig. \ref{single_rw} and characterized by its mass $m$, its axial and radial moment of inertia $J_w$ and $J_r$ respectively, its spinning rate $\Omega$ around its $\mathbf{z}_w$ axis, its center of mass $G$ and the connection point $P$ in the frame $\mathcal{R}_a = (O,\mathbf{x}_a,\mathbf{y}_a,\mathbf{z}_a)$. The input torque $-u$ is applied along the wheel axis by a driving mechanism.

The assumptions used to model the wheel are:

\begin{enumerate}
	\item The wheel mounting is assumed to be rigid in frame $\mathcal{R}_a$. This implies that $\mathbf{z}_w$ (the spin axis of the wheel) is constant in frame $\mathcal{R}_a$
	\item The wheel is considered balanced, that is the inertia tensor $\left[\mathbf{J}_{G}^\mathcal{RW}\right]_{\mathcal{R}_{w}}$ is diagonal
	\item The wheel frame $\mathcal{R}_w = (G,\mathbf{x}_w,\mathbf{y}_w,\mathbf{z}_w)$ is attached to the wheel axis and it is motionless in $\mathcal{R}_a$
	\item There is no torque disturbances about the wheel $\mathbf{z}_w$ (this assumption is not restrictive since the torque applied to the wheel $-u$ is an input of the model).
\end{enumerate}

To take into account the imbalances of the wheel responsible for harmonic disturbances \cite{preda:hal-01887323}, a vector of internal disturbances $\mathbf{w}_{rw}$, composed of the three components (in the wheel frame $\mathcal R_w$) of the disturbing force and two components of the radial disturbing torque, is considered acting at the center of mass of the wheel. The Newton-Euler equations applied to the RW at point $G$ reads:

\begin{equation}
	\begin{gathered}
		\mathbf{F}_{\mathcal{P} / \mathcal{RW}}+\mathbf{w}_{rw}\{1\}\mathbf{x}_{w}+\mathbf{w}_{rw}\{2\}\mathbf{y}_{w}+\mathbf{w}_{rw}\{3\}\mathbf{z}_{w}=m \mathbf{a}_{G} \\
		\mathbf{T}_{\mathcal{P} / \mathcal{RW}, G}+\mathbf{w}_{rw}\{4\} \mathbf{x}_{w}+\mathbf{w}_{rw}\{5\} \mathbf{y}_{w}=\left.\frac{d \mathbf{H}^{\mathcal{RW}}}{d t}\right|_{\mathcal{R}_{w}}+\boldsymbol{\omega} \times \mathbf{H}^{\mathcal{RW}}
	\end{gathered}
\end{equation}

where $\mathbf{H}^{\mathcal{RW}}=\mathbf{J}_{G}^{\mathcal{RW}} \boldsymbol{\omega}+J_{w} \Omega \mathbf{z}_{w}$ is the total angular momentum of the body $\mathcal{RW}$ and $\omega$ is the angular rate vector of the parent body with respect to the inertial frame.
The torque $\mathbf{T}_{\mathcal{P} / \mathcal{RW}, G}$ imposed by the spacecraft to the reaction wheel can be then rewritten as it follows:

\begin{equation}
	\mathbf{T}_{\mathcal{P} / \mathcal{RW}, G}+\mathbf{w}_{rw}\{4\} \mathbf{x}_{w}+\mathbf{w}_{rw}\{5\} \mathbf{y}_{w}=\mathbf{J}_{G}^{\mathcal{RW}} \dot{\boldsymbol{\omega}}+J_{w}\dot{\Omega} \mathbf{z}_{w}+J_{w} \Omega \left.\frac{d\left(\mathbf{z}_{w}\right)}{d t}\right|_{\mathcal{R}_{w}}+\boldsymbol{\omega} \times \mathbf{J}_{G}^{\mathcal{RW}}\bm{\omega}+\boldsymbol{\omega} \times J_{w} \Omega \mathbf{z}_{w}
\end{equation}

From the assumption that the spinning axis $\mathbf{z}_{w}$ stays constant in $\mathcal{R}_a$, the term $\left.\frac{d\left(\mathbf{z}_{w}\right)}{d t}\right|_{\mathcal{R}_{w}}$ is null. Discarding the second order term $\boldsymbol{\omega} \times \mathbf{J}_{G}^{\mathcal{RW}}\bm{\omega}$ and knowing that along $\mathbf{z}_{w}$:
\begin{itemize}
	\item $\mathbf{z}_{w}.\mathbf{T}_{\mathcal{P} / \mathcal{RW}, G} = -u$
	\item $-u = J_w \left(\dot{\Omega}+	\mathbf{z}_{w}.\dot{\bm{\omega}}\right)$,
\end{itemize}

the linear dynamic model of the wheel in the wheel frame $\mathcal{R}_w$, called $\left[\mathcal{M}^{\mathcal{RW}}_{G}\right]_{\mathcal{R}_w}(s)$ reads:

\begin{align}
	&\left[\mathbf{F}_{\mathcal{P} / \mathcal{RW}}\right]_{\mathcal{R}_w} =m \left[\mathbf{a}_{G}\right]_{\mathcal{R}_w}-\left[\mathbf{w}_{rw}\right]_{\mathcal{R}_w}\{1:3\}\\
	&\left[\mathbf{T}_{\mathcal{P} / \mathcal{RW}, G}\right]_{\mathcal{R}_w}\{1:2\}=\left(\left[\mathbf{J}_{G}^{\mathcal{RW}}\right]_{\mathcal{R}_w}-\dfrac{1}{s} J_{w} \Omega\left[\left({ }^{*} \mathbf{z}_{w}\right)\right]_{\mathcal{R}_w}\right)\{1:2,1:2\}\left[ \dot{\boldsymbol{\omega}}\right]_{\mathcal{R}_w}\{1:2\}-\left[\mathbf{w}_{rw}\right]_{\mathcal{R}_w}\{4:5\}\\
		&\left[\mathbf{T}_{\mathcal{P} / \mathcal{RW}, G}\right]_{\mathcal{R}_w}\{3\} = -u \\
	&\dot{\Omega}=-\dfrac{1}{J_{w}} u -\left[\dot{\bm{\omega}} \right]_{\mathcal{R}_w}\{3\}
\end{align}

\begin{figure}[h!]
	\centering
	\includegraphics[width=\columnwidth]{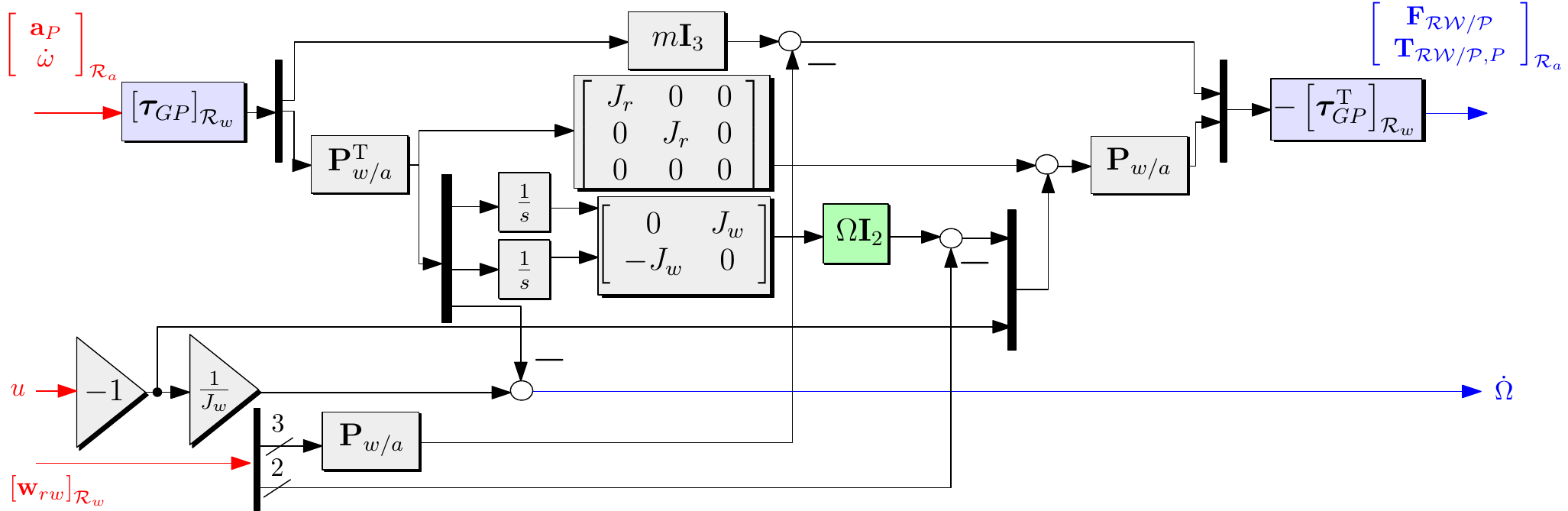}
	\caption{Reaction wheel TITOP model with internal disturbance $\left[\mathcal{M}^{\mathcal{RW}}_{P}\right]_{\mathcal{R}_a}(s)$ expressed in $\mathcal{R}_a$ w.r.t. attachment point $P$}
	\label{rwa_titop}
\end{figure}

\begin{figure}[htpb]
	\centering
		\includegraphics[scale=0.6]{RW_cluster.pdf}\,
\includegraphics[scale=0.5]{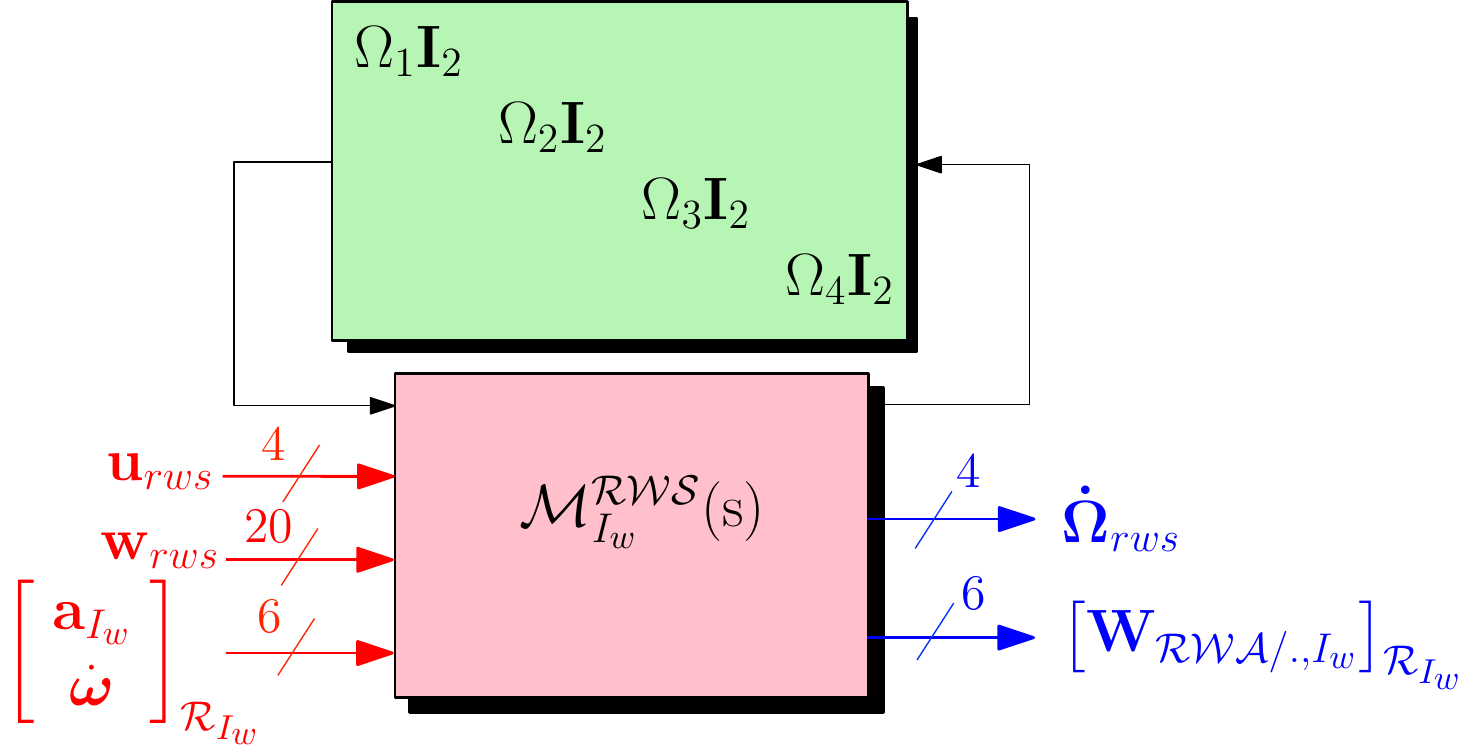}
	\caption{Equivalent LPV TITOP model of the RWS at anchorage point $I_w$ and scheduled according to the 4 RW speeds}	
	\label{eq_titop_rwa}
\end{figure}

Where $\left[\left({ }^{*} \mathbf{z}_{w}\right)\right]_{\mathcal{R}_{w}}=\left[\begin{array}{ccc}
	0 & -1 & 0 \\
	1 & 0 & 0 \\
	0 & 0 & 0
\end{array}\right]
$ is the skew-symmetric matrix associated with $\left[\mathbf{z}_{w}\right]_{\mathcal{R}_{w}}$.
Taking into account:
\begin{itemize}
	\item the direction cosine matrix (DCM) $\mathbf{P}_{w / a}$ from the frame $\mathcal{R}_a$ to the frame $\mathcal{R}_w$ such that:
	
	\begin{equation}
		\mathbf{P}_{w / a}=\mathbf{P}_{\mathbf{z} w}\left[\begin{array}{ccc}
			\operatorname{det}\left(\mathbf{P}_{\mathbf{z w}}\right) & 0 & 0 \\
			0 & 1 & 0 \\
			0 & 0 & 1
		\end{array}\right] \text { with: } \mathbf{P}_{\mathbf{z} w}=\left[\operatorname{null}\left(\left[\tilde{\mathbf{z}}_{w}\right]_{\mathcal{R}_{\mathfrak{a}}}^{T}\right) \quad\left[\tilde{\mathbf{z}}_{w}\right]_{\mathcal{R}_{a}}\right] \text { and } \tilde{\mathbf{z}}_{w}=\mathbf{z}_{w} /\left\|\mathbf{z}_{w}\right\|\,,
	\end{equation}
\item 	and  the kinematic model $\mathbf{\tau}_{GP}$ from point $G$ to point $P$ such that:

\begin{equation}
	\bm{\tau}_{GP} = 
	\left[\begin{array}{cc}
		\mathbf{I}_{3\times3} & \left({ }^{*} \overrightarrow{G P}\right) \\
		0_{3 \times 3} & \mathbf{I}_{3\times3}
	\end{array}\right]\,,
\end{equation}
\end{itemize}
the TITOP model $\left[\mathcal{M}^{\mathcal{RW}}_{P}\right]_{\mathcal{R}_a}(s)$ in $\mathcal{R}_a$ frame of a RW is then depicted in Fig. \ref{rwa_titop}. This model, parametrized according to the wheel speed $\Omega$, corresponds to a Linear Parameter Varying (LPV) system with the parameter $\Omega$ repeated twice.
This model has as input:
\begin{itemize}
	\item the acceleration vector $\ddot{\mathbf{x}}_p = \left[\mathbf{a}_P^\mathrm{T},\,\dot{\bm{\omega}}^\mathrm{T}\right]^\mathrm{T}$ of the anchor point $P$ of the RW to the parent structure expressed in the $\mathcal{R}_a$,
	\item the command torque $u$,
	\item the disturbance $\mathbf{w}_{rw}$.
\end{itemize}
The outputs of the model are:
\begin{itemize}
	\item the wrench of the forces/torques $\mathbf{W}_{\mathcal{RW/P},P}= \left[\mathbf{F}_{\mathcal{RW/P}}^\mathrm{T},\,\mathbf{T}_{\mathcal{RW/P},P}^\mathrm{T}\right]^\mathrm{T}$ imposed by the RW to the parent structure,
	\item the axial acceleration $\dot{\Omega}$ of the RW.
\end{itemize}
If a pyramidal reaction wheel system (RWS) is now considered to control the rigid motion of the spacecraft,
 the TITOP model of such assembly, called $\mathcal{M}^{\mathcal{RWS}}_{I_w}(s)$, is showcased in Fig. \ref{eq_titop_rwa} and easily built using 4 elementary blocks $\left[\mathcal{M}^{\mathcal{RW}}_{I_w}\right]_{\mathcal{R}_{I_w}}(s)$ expressed in $\mathcal{R}_{I_w}$, where $I_w$ is the anchor point of the RWS. Note that the kinematic model between each attachment point of the wheel and $I_w$ has been voluntary omitted for more readability of the Fig. \ref{eq_titop_rwa} (right). 

\subsection{Modeling of micro-vibration mitigation actuators}
\label{mv_act_sec}

In order to perform micro-vibration damping, a variety of control system architectures can be considered. Disturbances induced by internal mechanisms span a large frequency range and require broadband control ranging from a few tenth of \si{Hz} to typically 100 \si{Hz}. Pure passive isolation allow mitigation at high-frequencies (above 100 \si{Hz}) but reveals to be ineffective at lower frequencies. Hybrid architectures employing both passive and active mechanisms are proven to be effective across the whole spectrum of interest \cite{Boquet2013ActiveAP}.

\subsubsection{Fast Steering Mirror}
\label{sec:fsm}
\begin{figure}[h!]
	\centering
	\includegraphics[scale=0.6]{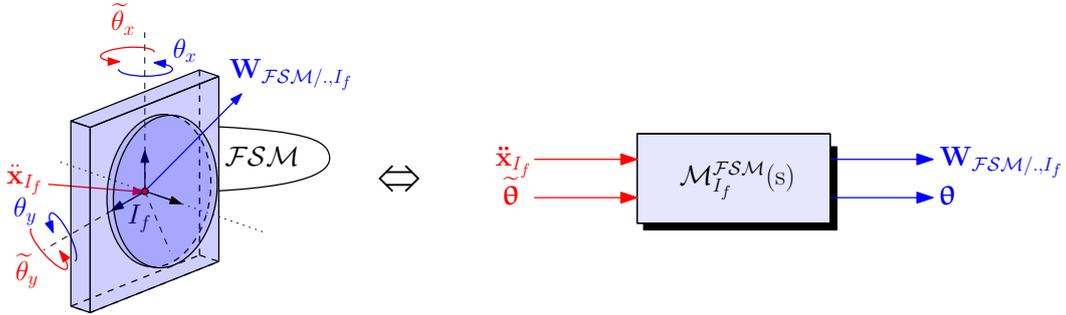}
	\caption{Piezoelectric fast steering mirror (left) and equivalent TITOP model (right)}
	\label{fsm}
\end{figure}

Science missions with optical payloads such as those at the core of space telescopes can benefit from active micro-vibration damping mechanisms directly located inside the payload as done in \cite{fsm_jwst}. Widely used actuators for this application are fast steering mirrors (FSM). These actuators are composed of a lightweight mirror attached to a steerable base that can rotate around two axes \cite{fsm_design} and are placed on the optical path of the payload, directly acting on the degraded LOS. Electro-magnetic FSMs (voice coils) and piezoelectric FSMs are the most widely used technologies nowadays \cite{SANFEDINO2022108168}. Voice coils can produce greater forces at moderate frequency and finer motion in comparison with piezoelectric FSMs. Those latter ones typically have the advantage of a wider bandwidth and a smaller volume. Another difference between these two technologies are their control signals: voice coils are driven by an input current translated into a force through the coil, whereas piezoelectric FSMs are driven by a modulated high voltage signal that directly results in small increments of motion. In this study, we will consider the piezoelectric FSM showcased in Fig. \ref{fsm}.

Such an actuator can be modeled as a rigid body $\mathcal{F}$ with two revolute joints $\mathcal{RJ}_x$ and $\mathcal{RJ}_y$ about its $\mathbf{x}$ and $\mathbf{y}$ axis and small stiffness and damping coefficients $k$ and $d$ at each joints. Each axis is directly controlled via angular position commands that relate with the input torques $T_x$ and $T_y$ through the joint stiffness. 

\begin{figure}[!h]
	\centering
	\includegraphics[width=.7\columnwidth]{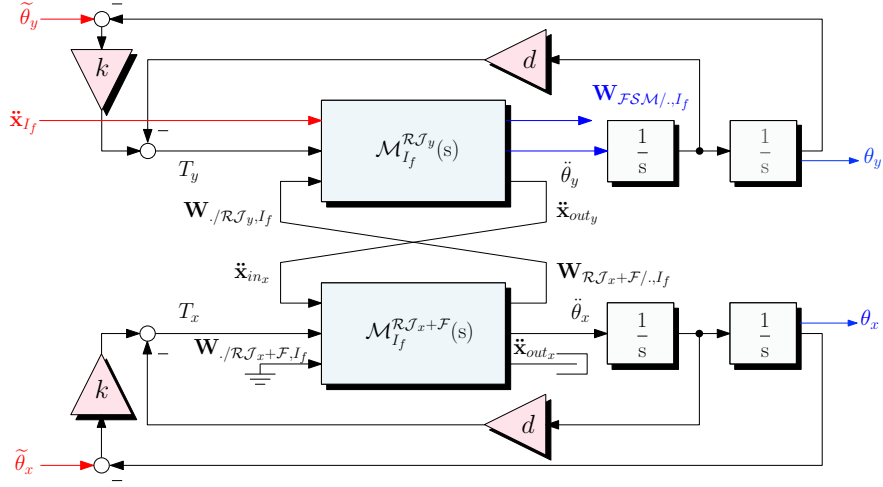}
	\caption{Piezoelectric FSM internal TITOP model}
	\label{fsm_titop_internal}
\end{figure}

The TITOP model of the piezoelectric FSM $\mathcal{M}_{I_f}^{\mathcal{FSM}}(s)$ anchored to the parent structure at the point $I_f$ is provided in Fig. \ref{fsm_titop_internal}. Note that $T_x$ and $T_y$ are the torques applied on the FSM about the $\mathbf{x}$ and $\mathbf{y}$ axis respectively and $\widetilde{\theta}_x$, $\widetilde{\theta}_y$ are the angular position input commands on the $\mathbf{x}$ and $\mathbf{y}$ axis.
The inputs of this model are:
\begin{itemize}
	\item the acceleration twist $\ddot{\mathbf{x}}_{I_f}$ of the anchor point $I_f$ to the parent structure,
	\item the two commanded angular positions $\tilde{\bm{\theta}}= \left[\tilde{\theta}_x,\,\tilde{\theta}_y\right]^{\mathrm{T}}$.
\end{itemize}
The outputs of the model are:
\begin{itemize}
	\item the wrench of the reaction forces/torques $\mathbf{W}_{\mathcal{FSM}/.,I_f}$ that the FSM imposes to the parent structure,
	\item the two actual FSM rotation angles ${\bm{\theta}}= \left[\theta_x,\,\theta_y\right]^{\mathrm{T}}$.
\end{itemize}

The models $\mathcal{M}^{\mathcal{RJ}_x}_{I_f}(\mathrm{s})$ and $\mathcal{M}^{\mathcal{RJ}_y+\mathcal{F}}_{I_f}(\mathrm{s})$ are respectively the TITOP models of a revolute joint around a given angular configuration and the TITOP models of a rigid body connected through a revolute joint at one port and loaded on the other port. These TITOP models are implemented in the SDT. For more details the reader is advised to refer to the SDT user-guide \cite{sdt_user_guide}.

\subsubsection{Proof-Mass  Actuators}
\label{sec:pma}
\begin{figure}[htpb]
	\centering
	\includegraphics[scale=0.6]{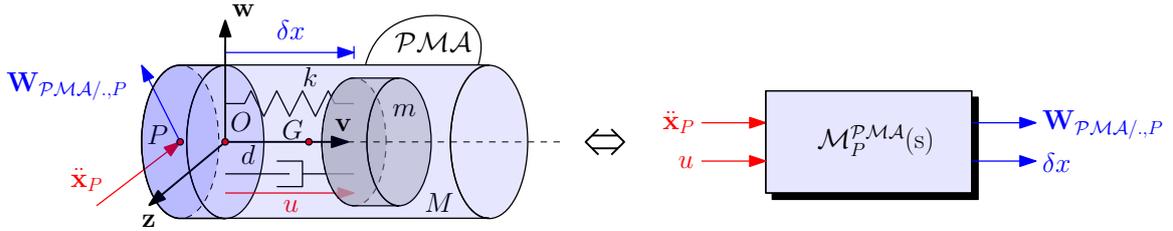}
	\caption{Proof-mass actuator view (left) and equivalent TITOP model (right)}
	\label{pma}
\end{figure}

Another possible micro-vibration mitigation architecture relies on proof-mass actuators (PMA). PMAs are mechanisms that can produce a periodic force along their principal axis. Multiple PMA can be placed and oriented in a particular configuration to provide controllability over the desired number of DOF of a body as in \cite{preda:tel-01722860}. Figure \ref{pma} shows a schematic view of a PMA and its equivalent TITOP model.

\begin{figure}[!h]
    \begin{subfigure}[b]{0.4\textwidth}
        \centering
	    \includegraphics[scale=0.75]{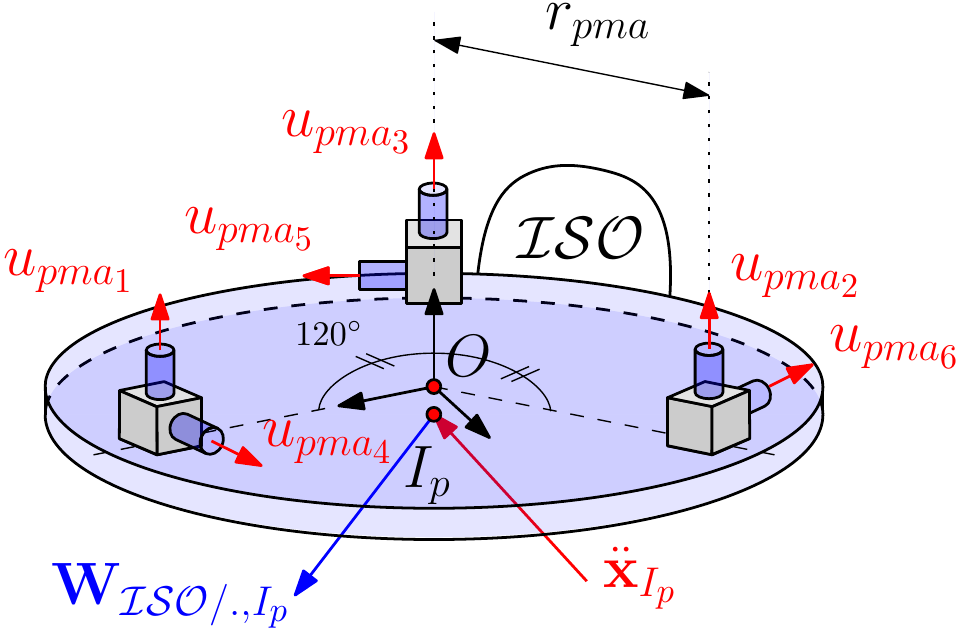}
	    \label{iso_pma}
    \end{subfigure}
    \hfill
    \begin{subfigure}[b]{0.6\textwidth}
        \centering
	    \includegraphics[scale=0.45]{iso_pma_titop.pdf}
	    \label{pma_iso_titop}
    \end{subfigure}
    \caption{Isolator and PMA assembly (left) and $\mathcal{M}^{\mathcal{IA}}_{I_p}(s)$ TITOP model (right)}
    \label{iso_pma_assembly}
\end{figure}

The moving mass $m$ of a single PMA is actuated by an input force $u$. The relative motion of the mass $m$ with respect to the PMA caging produces a reaction force on the parent body that allows damping vibrational motion. Points $O$, $G$, $P$ are respectively the reference point, the center of mass (COM) of the PMA at rest, and the connection point with the parent body, $\mathbf{J}_{G}^{PMA}$ is the total moment of inertia of the PMA, $m$ and $M$ are respectively the proof-mass and caging masses, $k$ and $d$ are respectively the proof-mass stiffness and damping, $\delta x$ is the relative displacement of the proof-mass with respect to the caging along the symmetry axis of the PMA, $\ddot{\mathbf{x}}_p$ is the acceleration twist of the PMA at the connection point $P$ expressed in $\mathcal{R}_a = \left(O,\mathbf{v},\mathbf{w},\mathbf{z}\right)$, $\mathbf{W}_{\mathcal{PMA}/.,P}$ is the wrench applied by the PMA at point $P$ expressed in $\mathcal{R}_a$. The linear dynamic equations governing the PMA, called $\mathcal{M}^{\mathcal{PMA}}_{P}(s)$, in the parent body (inherited) frame reads:

\begin{align}
	\label{eq_1_PMA}
	m\left(
	\begin{bmatrix}
		\mathbf{v}^{T} & \mathbf{0}_{1\times3}
	\end{bmatrix}
	\boldsymbol{\tau}_{G P} \ddot{\mathbf{x}}_{P}+{\delta}\ddot x\right)&=-k \delta x-d {\delta}\dot x+u \\
	\label{eq_2_PMA}
	\mathbf{W}_{\mathcal{PMA} / ., P}&=-\boldsymbol{\tau}_{G P}^{T}\left(\left[\begin{array}{cc}
		M \mathbf{I}_{3} & \mathbf{0}_{3\times3} \\
		\mathbf{0}_{3\times3} & \mathbf{J}_{G}^{\mathcal{PMA}}
	\end{array}\right] \boldsymbol{\tau}_{G P} \ddot{\mathbf{x}}_{P}+m\left[\begin{array}{l}
		\mathbf{v} \\
		\mathbf{0}_{3\times3}
	\end{array}\right] {\delta}\ddot x\right)
\end{align}
The equivalent TITOP model of the PMA is shown in Fig. \ref{pma} (right), where the inputs are:
\begin{itemize}
	\item  the acceleration vector $\ddot{\mathbf{x}}_P$ of the anchor point $P$ to the parent structure,
	\item the input force $u$.
\end{itemize}
The outputs of this models are:
\begin{itemize}
	\item the wrench of the reaction forces/torques $\mathbf{W}_{\mathcal{PMA}/.,P}$ that the FSM imposes to the parent structure,
	\item the relative displacement $\delta x$ of the proof-mass.
\end{itemize}
Multiple PMAs can be mounted on a passive isolator to provide wide-band micro-vibration attenuation \cite{preda:tel-01722860} as showcased in Fig. \ref{iso_pma_assembly}. This actuator placement has been chosen in such a way that each of the 6 DOFs of the isolator is controllable. The TITOP model of the isolator assembly (IA), called $\mathcal{M}^{\mathcal{IA}}_{I_P}(s)$, is composed of a 6-DOF mass-spring-damper system $\mathcal{M}_{I_P}^{\mathcal{ISO}}(s)$ (see \cite{sdt_user_guide} for more details) to model the isolator and 6 occurrences of $\mathcal{M}^{\mathcal{PMA}}_{P}(s)$.
Note that, for more readability of the figure, the kinematic model between each PMA's connection point and the central point $O$ has been voluntary omitted.

\subsection{Full spacecraft assembly}
\label{sc_dyn_sec}

\begin{figure}[h!]
	\centering
	\includegraphics[scale=0.45]{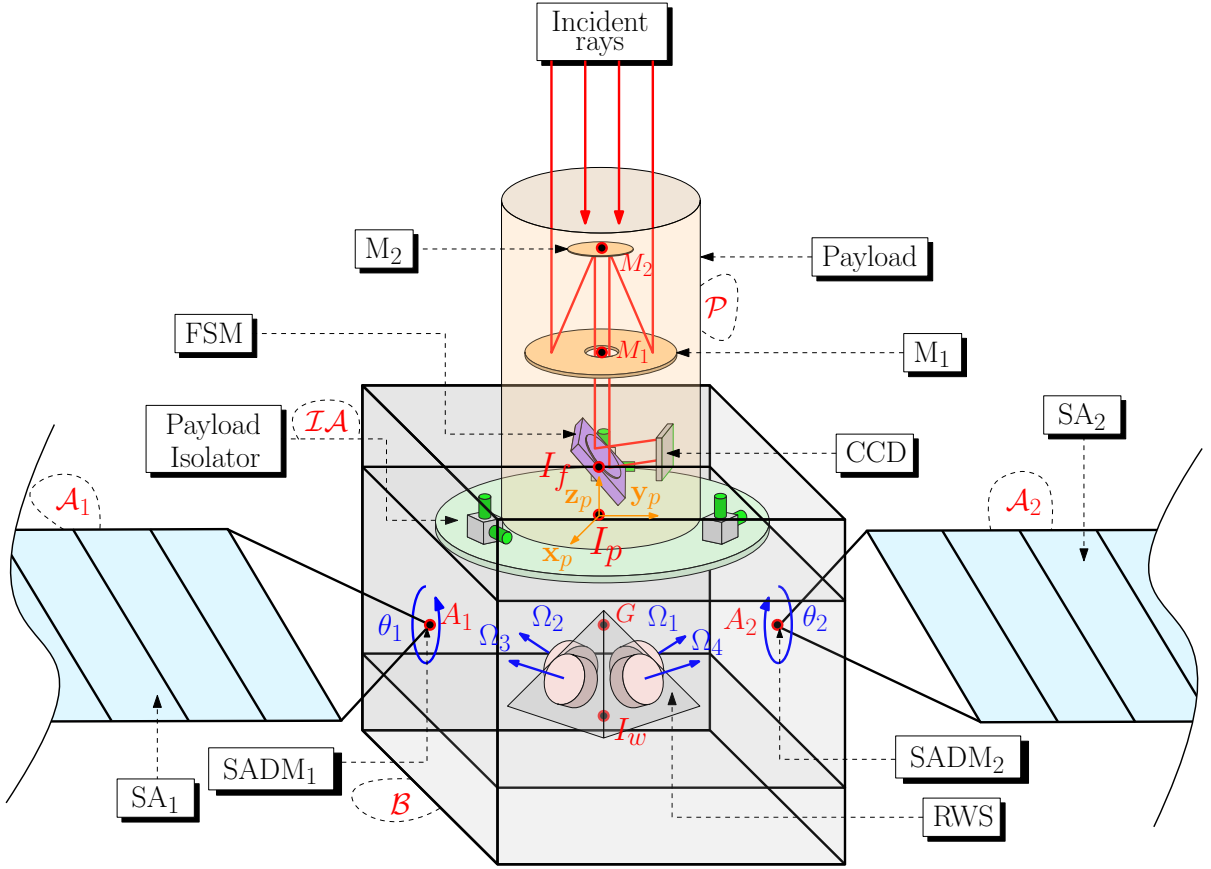}
	\caption{Spacecraft architecture with essential nomenclature}
	\label{fig:full_sc}
\end{figure}

In a spirit of comparing multiple micro-vibration control system architectures, this paper proposes to assess their respective performances (disturbance rejection, stability, robustness) on a dummy space telescope from a generic observation mission. The satellite dynamics can be assembled with the TITOP models derived in the previous sections, in a MATLAB/Simulink environment and the SDT. The proposed spacecraft in Fig. \ref{fig:full_sc} is composed of a central flexible body $\mathcal{B}$ connected to two rotating solar arrays, $\mathcal{A}_1$ and $\mathcal{A}_2$, at points $A_1$ and $A_2$ respectively, one flexible optical payload $\mathcal{P}$ and one RWS, respectively connected at two distinct points $I_p$ and $I_w$. Moreover the payload is anchored at the spacecraft through an isolator assembly as the one presented in section \ref{sec:pma} at point $I_p$. The optical payload is composed of a flexible structure enclosing the optical elements: the two mirrors $M_1$ and $M_2$, the charge-coupled device (CCD) and the FSM. The physical parameters of the model, i.e. flexible modes and dampings, are input data from an industrial benchmark, imported in the SDT directly with PATRAN/NASTRAN files \cite{sdt_user_guide}.

\begin{figure}[h!]
	\centering
	\includegraphics[scale=0.475]{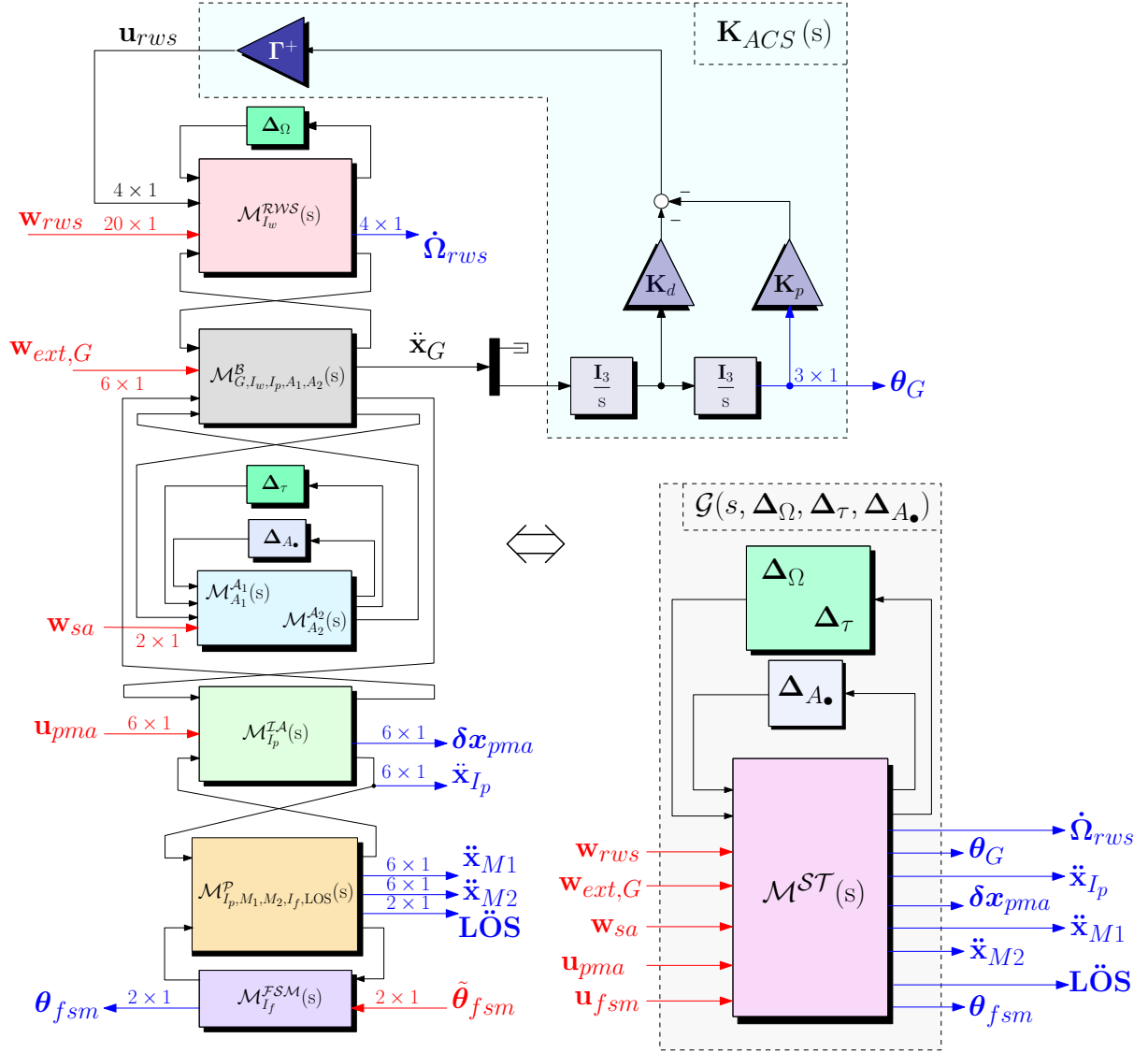}
	\caption{Full space telescope TITOP interconnection diagram (left) and equivalent LFT model (right)}
	\label{fig:full_sc_int_TITOP}
\end{figure}

Thanks to SDT, it is possible to build the entire spacecraft model by interconnections of elementary blocks corresponding to each sub-structure as depicted in Fig. \ref{fig:full_sc_int_TITOP}. Note that blocks' colors are the same as each sub-system in Fig. \ref{fig:full_sc}. 
For more details on how to connect several blocks in SDT please refer to \cite{sanfe2019}.

In Fig. \ref{fig:full_sc_int_TITOP} several models are found:
\begin{itemize}
	\item $\mathcal{M}_{G,I_w,I_p,A_1,A_2}^{\mathcal{B}}(s)$ is the TITOP model of the flexible central body $\mathcal{B}$ imported from NASTRAN with parent point $G$ (center of mass) and children connection points $I_w$, $I_p$, $A_1$, $A_2$. The external wrench $\mathbf{w}_{ext,G}$ acting at $G$ is the input of this model and the acceleration twist $\ddot{\mathbf{x}}_G$ of the point $G$ is the output;
	\item $\mathcal{M}_{A_1}^{\mathcal{A}_1}(s)$ and $\mathcal{M}_{A_2}^{\mathcal{A}_2}(s)$ are respectively the TITOP models of the two flexible solar panels $\mathcal{A}_1$ and $\mathcal{A}_2$, connected to $\mathcal{B}$ at $A_1$ and $A_2$ respectively through two identical SADM, taking into account the stiffness of a reduction gearbox. See \cite{SANFEDINO2022108168} for more detail on SADM TITOP model. The block $\bm{\Delta}_\tau=\tau\mathbf{I}_{32}$ is the parametric uncertainty of the two solar arrays configurations modeling the different orientation of the rotating SA where $\tau=\tan(\theta/4)$ is the parametrization of the SA rotation angle $\theta$ presented in \cite{SANFEDINO2022108168}. The uncertain block $\bm{\Delta}_{A_\bullet}$ takes into account the uncertainties on the first two frequencies of the flexible modes of the two SA. $\mathbf{w}_{sa}$ is the vector of the $2$ disturbance torques transmitted by the two SADM driving signals to the SA rotation axes;
	\item $\mathcal{M}_{I_w}^{\mathcal{RWS}}(s)$ is the model of the assembly of four reaction wheels presented in section \ref{mv_sources_sec}. The block $\bm{\Delta}_\Omega=\operatorname{blkdiag}\left(\Omega_1\mathbf{I}_2,\Omega_2\mathbf{I}_2,\Omega_3\mathbf{I}_2,\Omega_4\mathbf{I}_2\right)$ takes into account the four spin rates as varying parameters. The harmonic disturbance vector induced by the four RWs is taken into account in the input $\mathbf{w}_{rws}$;
	\item $\mathcal{M}_{I_p,M_1,M_2,I_f,\mathrm{LOS}}^{\mathcal{P}}(s)$ is the TITOP model of the flexible payload $\mathcal{P}$ imported from NASTRAN with parent point $I_p$ (connection with $\mathcal{B}$) and children connection points $M_1$ (connection point with mirror $\mathcal{M}_1$), $M_2$ (connection point with mirror $\mathcal{M}_2$), $I_f$ (connection point with payload isolator $\mathcal{ISO}$), $\mathrm{LOS}$ (connection point with the CCD camera). $\mathcal{M}_{I_p,M_1,M_2,I_f,\mathrm{LOS}}^{\mathcal{P}}(s)$ outputs the acceleration vectors $\ddot{x}_{M_1}$ and $\ddot{x}_{M_2}$ of points $M_1$ and $M_2$ respectively and the two angular accelerations $\ddot{{\textbf{LOS}}}$ of the LOS, whose double integration is measured by the CCD camera;
	\item $\mathcal{M}^{\mathcal{IA}}_{P,I}(s)$ is the isolator and PMA asembly  showcased in Fig. \ref{iso_pma_assembly}. The 6 PMA control signals $\mathbf{u}_{pma}$ are inputs to this block. The outputs are the acceleration vector $\ddot{\mathbf{x}}_{I_p}$ of point $I_p$ and the six relative displacements $\bm{\delta x}_{pma}$ of the 6 PMAs;
	\item $\mathcal{M}_{I_f}^{\mathcal{FSM}}(s)$ is the TITOP model of the FSM presented in section \ref{sec:fsm}. The inputs to this model are the tip/tilt FSM commanded angles $\tilde{\bm{\theta}}_{fsm}$ and the outputs are the actual tip/tilt FSM angles ${\bm{\theta}}_{fsm}$.	
\end{itemize}

In order to control the spacecraft attitude, a proportional-derivative (PD) controller tuned on the total inertia matrix (assumed uncoupled) with respect to the central body center of gravity $G$, $\mathbf{J}^{\text{tot}}=\operatorname{blkdiag}\left(\mathbf{J}_{x}^{\text {tot }}, \mathbf{J}_{y}^{\text {tot }}, \mathbf{J}_{z}^{\text {tot }}\right)$, is proposed:

\begin{equation}
	\mathbf{u}_{rws}=-\bm{\Gamma}^{+}\left(\mathbf{K}_{d} \dot{\boldsymbol{\theta}}_{G}-\mathbf{K}_{p} \boldsymbol{\theta}_{G}\right),
\end{equation}
where $\mathbf{K}_{p} =\omega_{\mathrm{ACS}}^{2} \mathbf{J}^{\text{tot}}, \hspace{1mm}$ $\mathbf{K}_{d}=2 \zeta_{\mathrm{ACS}} \omega_{\mathrm{ACS}} \mathbf{J}^{\text{tot}}$ (with $\omega_{\mathrm{ACS}} = 0.06$ rad/s, $\zeta_{\mathrm{ACS}} = 0.7$). $\bm{\Gamma}^{+}$ is the Moore-Penrose pseudo-inverse of the $3\times 4$ RWS spin axis matrix $\bm{\Gamma}$ whose columns are the 4 wheels spin axes expressed in the parent (inherited) frame. The total inertia matrix $\mathbf{J}^{\text{tot}}$ can be obtained by computing the inverse low-frequency (DC) gain of the transfer from the body torque disturbances $\mathbf{w}_{ext,G}\{4:6\}$ to the body angular acceleration $\bm{\theta_{G}}=\mathbf{\ddot{x}}_G\{4:6\}$. 
As shown in Fig. \ref{fig:full_sc_int_TITOP} the assembled model of the space telescope built in SDT is equivalent to an LFT model $\mathcal{G}(s,\bm{\Delta}_\Omega, \bm{\Delta}_\tau,\bm{\Delta}_{A_\bullet})=\mathcal{F}_u(\mathcal{M}^{ST}(s),\mathrm{diag}(\bm{\Delta}_\Omega, \bm{\Delta}_\tau,\bm{\Delta}_{A_\bullet}))$. $\mathcal{G}(s,\bm{\Delta}_\Omega, \bm{\Delta}_\tau,\bm{\Delta}_{A_\bullet})$
is an uncertain minimal state-space model of order 204 with 2 occurrences of $\Omega_1$, 2 occurrences of $\Omega_2$, 2 occurrences of $\Omega_3$, 2 occurrences of $\Omega_4$, 4 occurrences of $\omega^{\mathcal{A}_\bullet}_1$, 4 occurrences of $\omega^{\mathcal{A}_\bullet}_2$ and 32 occurrences of $\tau$.

\subsection{Analysis of RW and SADM perturbation on the space telescope LOS}

\begin{figure}[!h]
	\centering
	\includegraphics[width=.85\columnwidth]{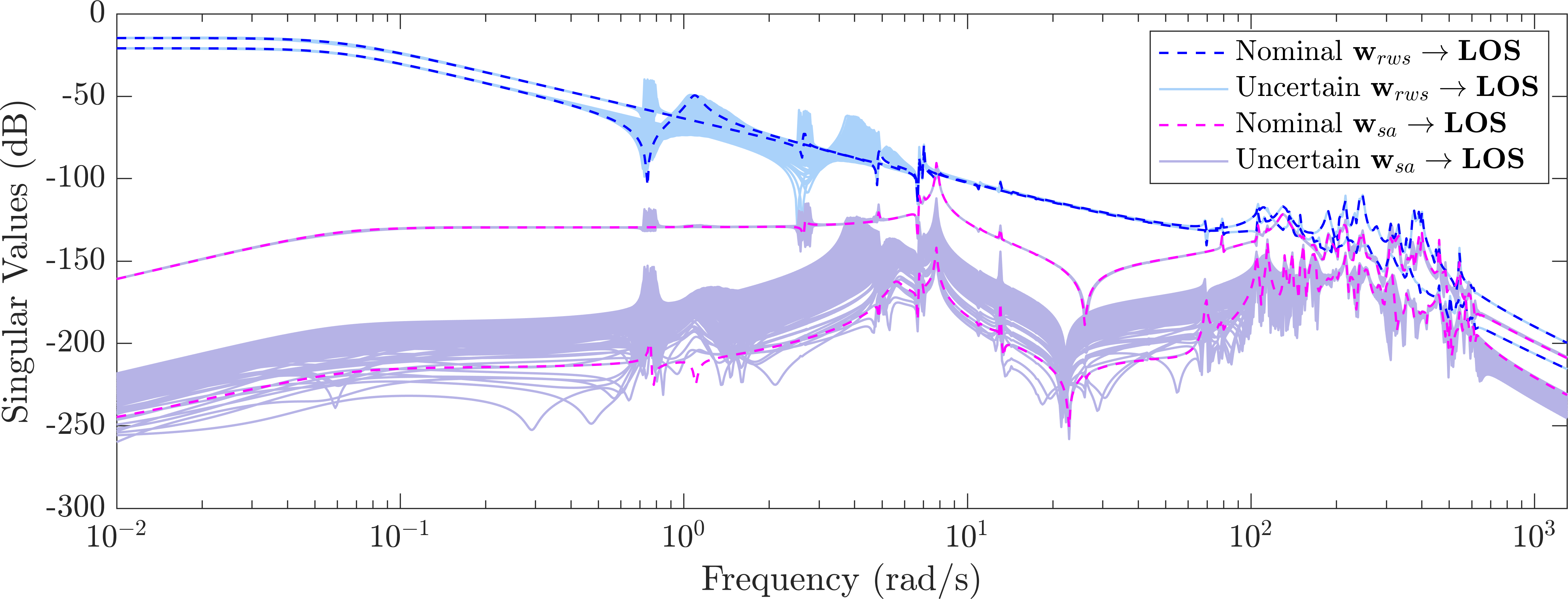}
	\caption{Transmissibility from microvibration disturbance sources (combination of RW and SADM) to the LOS for the plant $\mathcal{G}(s,\bm{\Delta}_\Omega, \bm{\Delta}_\tau,\bm{\Delta}_{A_\bullet})$}
	\label{fig:dist2LOS}
\end{figure}

Once the system interconnection is available as shown in section \ref{sc_dyn_sec}, a complete understanding of the physics of the system is possible. In particular, it is interesting to analyze how the microvibration perturbations, coming from RW imbalances and SADM driving torques, propagate trough the whole spacecraft flexible structure to the CCD camera and impact the LOS. 
As an example Fig. \ref{fig:dist2LOS} shows the singular values of both the transfer functions from all reaction wheel harmonic perturbation signals $\mathbf{w}_{rws}$ to the LOS error and from the SADM driving torque $\mathbf{w}_{sa}$ to the LOS error.
On the transfer from $\mathbf{w}_{sa}$ to the LOS error, the high-pass response visible at low frequency ($[0.01,\; 1]$ rad/s) is due to the SADM gearbox stiffness.
At low frequency $[0,\;0.05]\,\mathrm{rad/s}$, the RW disturbance $\mathbf{w}_{rws}$ is filtered by the ACS.  
From the Fig. \ref{fig:dist2LOS}, between $1$ and $100\,rad/s$, it can be noticed that the reaction wheel disturbance can propagate through the structure way higher (up to 3 order of magnitude) than their correspondent SADM perturbation. For higher frequencies greater than $100\,rad/s$ the two kinds of perturbation are comparable. 

\begin{figure}[!h]
	\centering
	\begin{subfigure}[b]{0.49\columnwidth}
		\centering
		\includegraphics[width=\columnwidth]{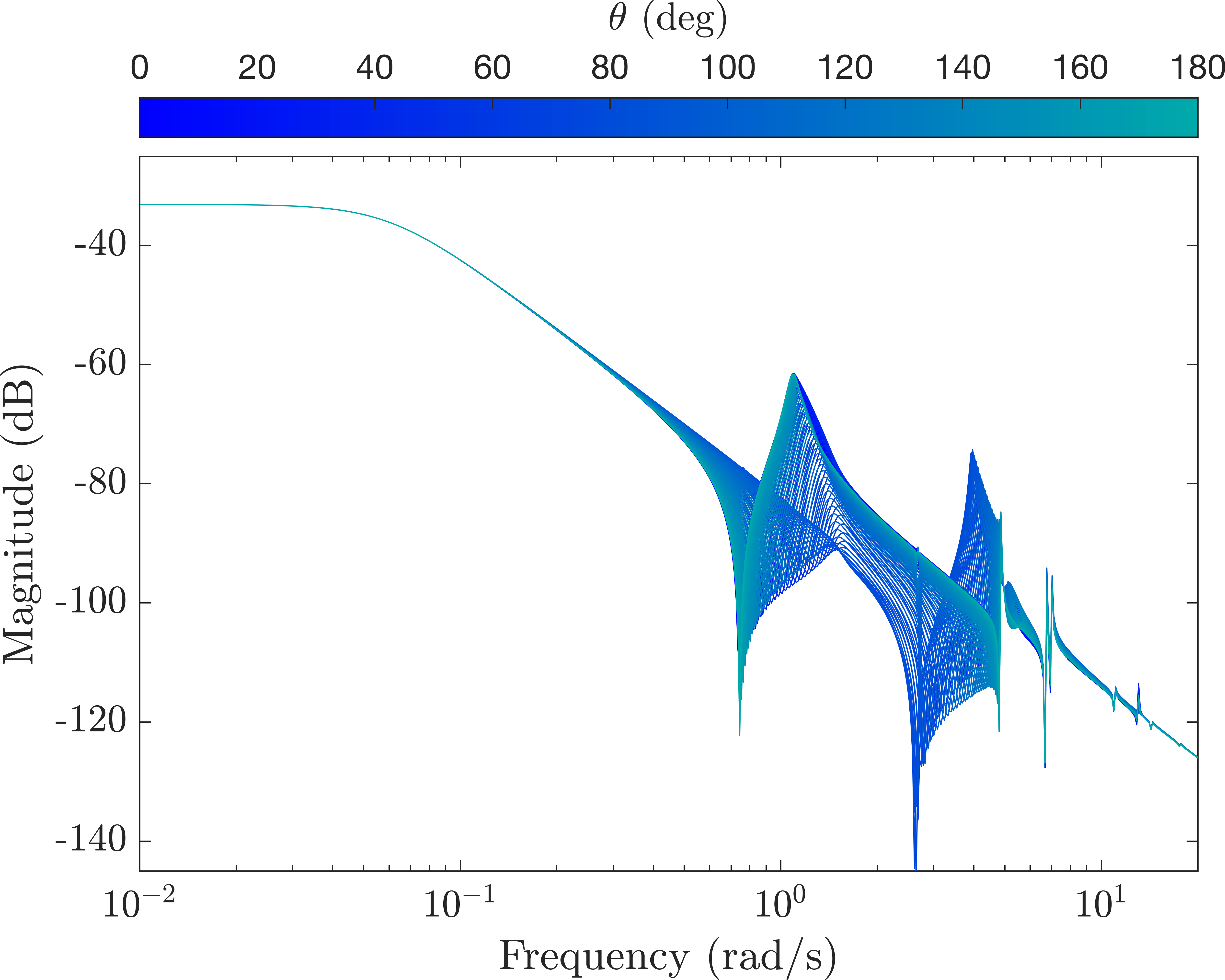}
		\caption{}
		\label{fig:tau_variation}
	\end{subfigure}
	\begin{subfigure}[b]{0.49\columnwidth}
		\centering
		\includegraphics[width=\columnwidth]{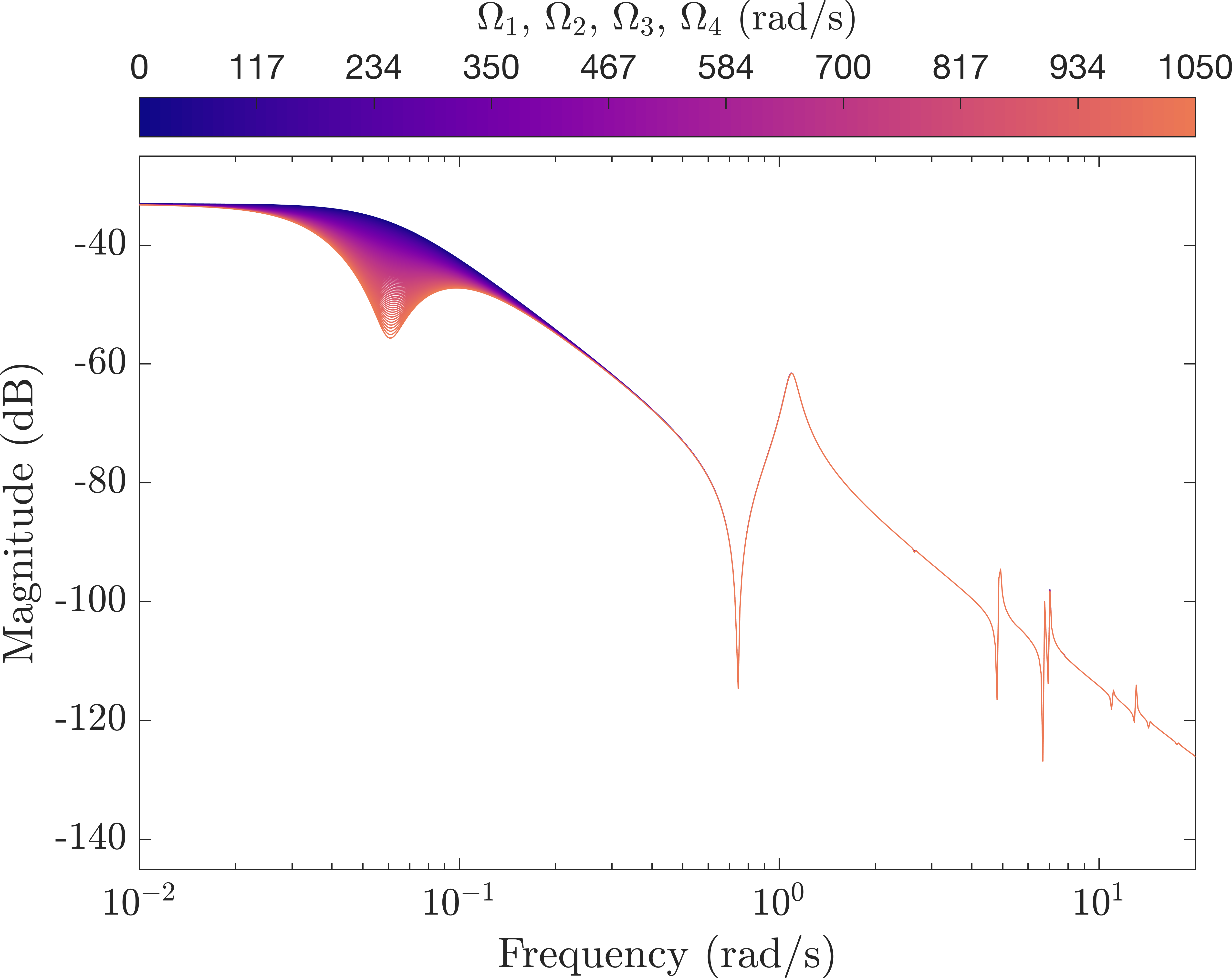}
				\caption{}
		\label{fig:omega_variation}
	\end{subfigure}
	\caption{Transfer function between $\mathbf{w}_{rws}\{1\}$ and $\textbf{LOS}\{1\}$ of the plant $\mathcal{G}(s,\bm{\Delta}_\Omega, \bm{\Delta}_\tau,\bm{\Delta}_{A_\bullet})$, (a) when the two solar panels rotate and the other varying and uncertain parameters have their nominal values; (b) when the four reaction wheels accelerate from 0 to their maximum speed at the same time and the other varying and uncertain parameters have their nominal values}
	\label{fig:variation}
\end{figure}

Another phenomenon to be highlighted in this complex scenario is the shifting of the spacecraft flexible modes due to both the rotation of the solar panels and to the speeds of the four reaction wheels. Figure \ref{fig:variation} depicts this behavior by analyzing the transfer function between the first reaction wheel harmonic disturbance $\mathbf{w}_{rws}\{1\}$ and the first LOS error $\textbf{LOS}\{1\}$. In particular Fig. \ref{fig:tau_variation} shows the evolution of system $\mathcal{G}(s,\bm{\Delta}_\Omega, \bm{\Delta}_\tau,\bm{\Delta}_{A_\bullet})$ when the two solar arrays driven by their SADM rotate and all the other uncertain/varying parameters are fixed to their nominal values. We notice that the coupling of the SA with the rest of the structure determines a shift of the first two solar array flexible modes according to their angular configuration. This means that if the frequency of the RW harmonic disturbance, linked to the wheel speed, falls exactly in correspondence of one of this shifting structural modes, a degradation of the LOS has to be expected. 
The rotation of the wheels influences also the transfer $\mathbf{w}_{rws}\{1\}\rightarrow \textbf{LOS}\{1\}$ as shown in Fig. \ref{fig:omega_variation} at very low frequencies. 
These two phenomena show how it is important to know in an early phase of a project the evolution of the system and its uncertainties in order to synthesize a robust control law in spite of all kind of predicted performance degradation. The way to model a complex system in a parametric-dependent multi-body framework as proposed in this paper perfectly copes with this need by integrating all the possible fluctuations of the plant in a unique LFT model. 

A controller designed on a model ignoring these varying/uncertain parameters could bring to discard the overall design during validation phase. Both conservatism (due to large margins considered in control design to compensate for possible uncertainties) and time consuming validation campaigns can then be avoided thanks to the proposed modeling approach.

\section{Fine LOS control}
\label{fine_los_control}

\begin{figure}[!h]
	\centering
	\includegraphics[width=.85\columnwidth]{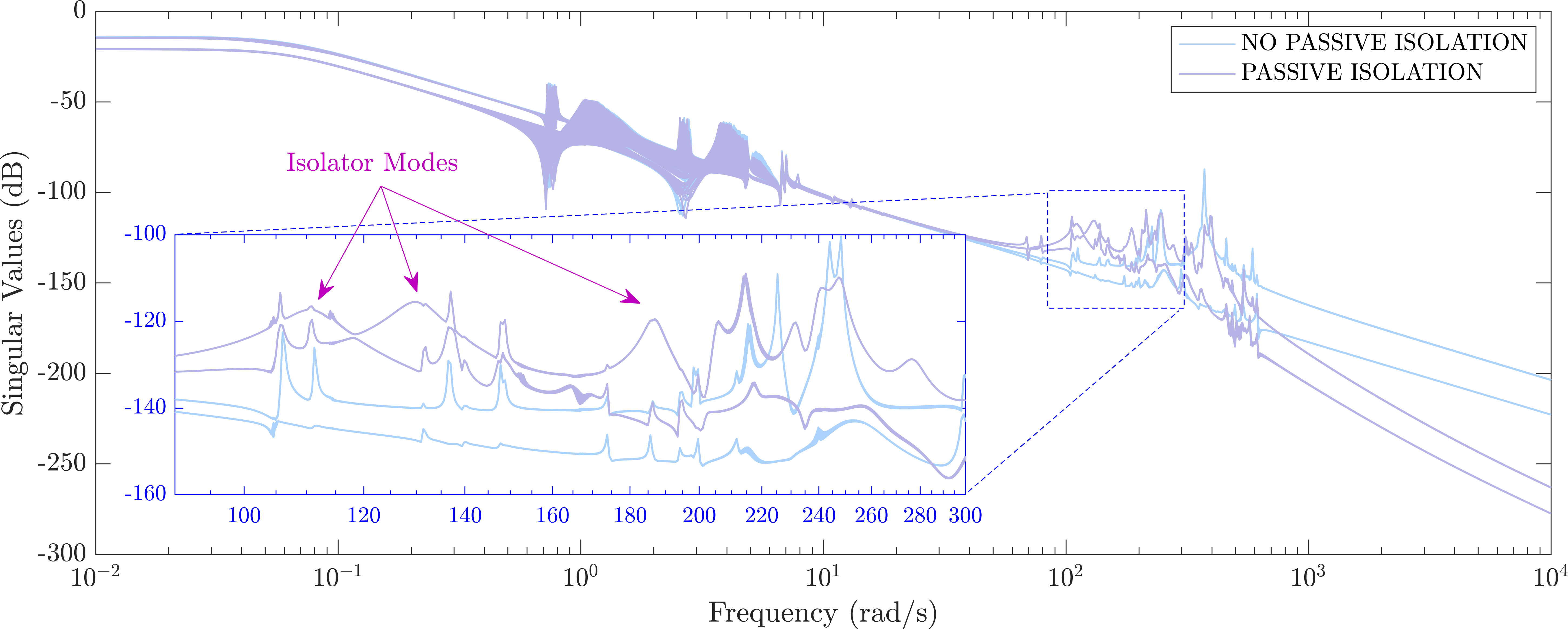}
	\caption{Transmissibility from microvibration disturbance sources (both RW and SADM) to the LOS for the plant $\mathcal{G}(s,\bm{\Delta}_\Omega, \bm{\Delta}_\tau,\bm{\Delta}_{A_\bullet})$ without and with a payload passive isolator}
	\label{fig:passiveVsOL} 
\end{figure}

When dealing with microvibrations, the first way to counteract their influence at very high frequency is to use passive isolation. If this strategy allows having an equivalent low-pass filter behavior for high frequencies, on the other hand it introduces some supplementary flexible modes at lower frequencies as shown in Fig. \ref{fig:passiveVsOL}.
For this reason, an hybrid (passive + active) control strategy is needed to mitigate microvibrations in the middle range frequencies. 
Two complementary active control architectures are presented in sections \ref{sec:fsm_control} and \ref{sec:FEM+PMA}.

\subsection{Hybrid control of LOS with FSM}
\label{sec:fsm_control}

A piezo-electric FSM is positioned on the optical path of the payload to perform active control at LOS level. Such actuator can perform control over a wide bandwidth, typically ranging from a few Hz to a few hundreds of Hz, which makes it efficient over the entire active control bandwidth. The FSM being controlled in position, one can directly cancel the motion of the LOS by applying the negative of the LOS to the FSM input. The main drawback of such architecture is the need to estimate the LOS at higher frequency than the one offered by direct measurements of a CCD camera $\textbf{LOS}_c^m$, that does not generally overcome few tens of Hz. For this reason, an estimation of the LOS is done by blending the camera measurements with the measurements provided by the accelerometers (at payload isolator $\ddot{\mathbf{x}}_{I_p}^m$ and at mirrors $M_1$ and $M_2$, respectively $\ddot{\mathbf{x}}_{M1}^m$ and $\ddot{\mathbf{x}}_{M2}^m$) and the FSM strain gauges (measuring the FSM tip/tilt deflections $\bm{\theta}_{fsm}^m$) along the optical path till the CCD camera. The FSM control law is thus an observer-based controller that reads:

\begin{equation}
	\mathbf{u}_{fsm} = -\mathbf{S}^{-1}_{FSM}\widehat{\mathbf{LOS}}
\end{equation}

Where $\mathbf{S}_{FSM} = \text{diag}\left(0.1,0.1\right)$ is the FSM sensitivity matrix that relates the motion of the FSM to the LOS, and $\widehat{\mathbf{LOS}}$ is an estimate of the LOS. 

\begin{figure}[!h]
	\centering
	\includegraphics[width=.8\columnwidth]{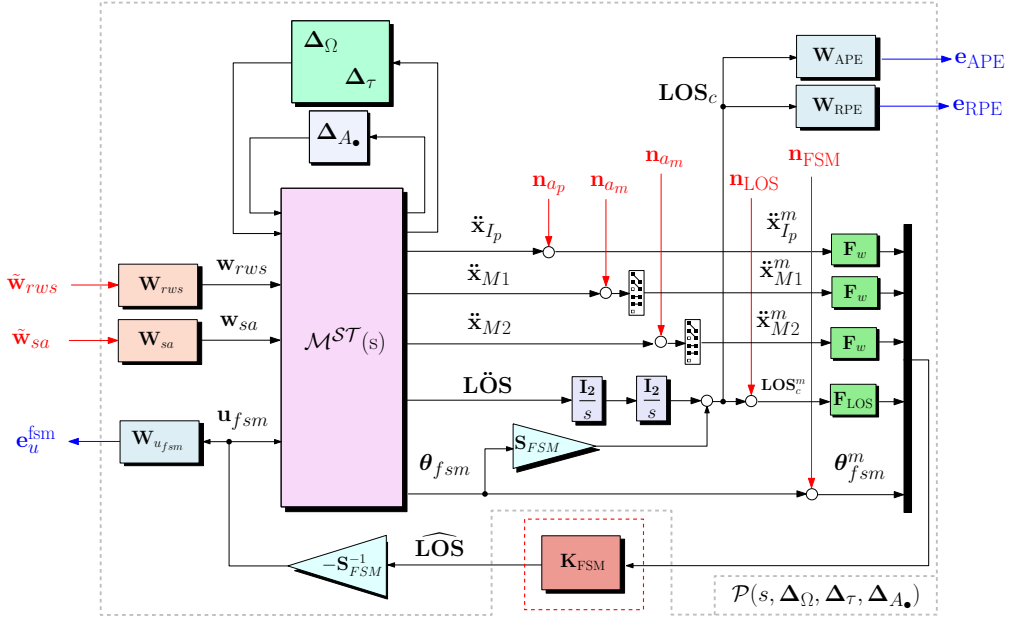}
	\caption{Control architecture for FSM robust control synthesis}
	\label{fig:synthesis_FSM} 
\end{figure}

The control architecture for FSM robust control synthesis is shown in Fig. \ref{fig:synthesis_FSM}, where the generalized plant $\mathcal{P}(s,\bm{\Delta_\Omega},\bm{\Delta_\tau},\bm{\Delta_{A_\bullet}})$ with normalized input/output weighting filters and the FSM controller $\mathbf{K}_{FSM}(\mathrm s)$ are depicted. The objective is to obtain the $2\times 18$ dynamic observer $\mathbf{K}_{FSM}(\mathrm s)$ by ensuring a prescribed level of pointing performance (LOS error) given an expected amplitude of microvibration disturbance as input to the system and coping with FSM actuation authority and all identified model uncertainties and varying parameters. Note that in the diagram a washout filter $\mathbf{F}_w(\mathrm s)$ is applied to the accelerometer measurements to reject sensor bias:
\begin{equation}
	\mathbf{F}_w(\mathrm s) = \frac{\mathrm s}{\mathrm s + 0.1}\textbf I_6
\end{equation} 
In order to take into account the low-pass behavior of the CCD camera, the filter $\mathbf{F}_\mathrm{LOS}$ is used as well:
\begin{equation}
	\mathbf{F}_\mathrm{LOS}(\mathrm s) = \frac{100}{\mathrm s^2 + 14\mathrm s + 100}\textbf{I}_2
\end{equation}
The set of sensors is characterized by the following levels of white Gaussian noise,  with their standard deviations $\bm{\sigma}$ and sampling times $dt$:
\begin{itemize}
	\item accelerometers noise $\mathbf n_{a_p}$, $\mathbf n_{a_m}$: $\bm{\sigma}_a = \mathrm{blkdiag}\left(0.0012\,\textbf{I}_3\, (\mathrm{m/s^2/\sqrt{Hz}}),\,0.0023\,\textbf{I}_3\, (\mathrm{rad/s^2/\sqrt{Hz}})\right)$ , $dt_a = 1\,\mathrm{ms}$;
	\item CCD noise $\mathbf n_{\mathrm{LOS}}$: $\bm{\sigma}_{\mathrm{LOS}} = 10^{-8}\,\textbf{I}_3\,(\mathrm{rad/\sqrt{Hz}})$, $dt_{\mathrm{LOS}} = 1\,(\mathrm{ms})$;
	\item strain gauge noise $\mathbf n_\mathrm{FSM}$: 
	$\bm{\sigma}_{\mathrm{FSM}} = 10^{-8}\,\textbf{I}_3\,(\mathrm{rad/\sqrt{Hz}})$, $dt_{\mathrm{FSM}} = 1\,\mathrm{ms}$
\end{itemize}

For the $\mathcal{H}_\infty$ robust control synthesis, we specify the following input weighting functions:
\begin{itemize}
	\item $\mathbf{W}_{rws}$ shapes the amplitude of the expected five harmonic perturbations for each of the four RW. For the $i$-th wheel this filter takes the form:
	\begin{equation}
		\mathbf{W}_{rws}^i(\mathrm s) = \mathrm{diag}(0.4\,\mathrm{N},\, 0.4\,\mathrm{N},\,0.35\,\mathrm{N},\, 0.3\,\mathrm{Nm},\,0.3\,\mathrm{Nm})\cdot\frac{\mathrm s+5.101\cdot 10^{-5}}{\mathrm s+5.101}\textbf{I}_5
	\end{equation} 
\item $\mathbf{W}_{sa}$ fixes the upper bound of the SADM input disturbance torque:
\begin{equation}
	\mathbf{W}_{sa} = 0.1\,\mathrm{Nm}\cdot \textbf{I}_2
\end{equation}
\end{itemize}
On the other hand, the following output weighting functions have been considered:
\begin{itemize}
	\item $\mathbf{W}_\mathrm{APE}$ bounds the maximum tolerated Absolute Performance Error (APE) \cite{Ott} on the LOS:
	\begin{equation}
		\mathbf{W}_\mathrm{APE} = \epsilon_\mathrm{APE}^{-1}\textbf{I}_2
	\end{equation}
	In this study $\epsilon_\mathrm{APE} = 10\,\mathrm{arcsec}$.
	\item $\mathbf{W}_\mathrm{RPE}$ defines the expected pointing performance in terms of Relative Performance Error (RPE), which is defined in \cite{Ott} as the  angular difference between the instantaneous LOS error vector  and the short-time average LOS error vector during a given integration time period $t_\Delta$. In frequency domain the RPE performance corresponds to the high-pass performance weight applied to the LOS error signal:
	
	\begin{equation}
		\mathbf{W}_\mathrm{RPE}(\mathrm s) = \epsilon_\mathrm{RPE}^{-1} \frac{t_\Delta \mathrm s\left(t_\Delta \mathrm s + \sqrt{12}\right)}{\left(t_\Delta \mathrm s\right)^2+6\left(t_\Delta \mathrm s\right)+ 12}\textbf{I}_2
	\end{equation}
where $\epsilon_\mathrm{RPE}$ overbounds the maximum RPE target value. For the present study case $t_\Delta = 20\,\mathrm{ms}$ and $\epsilon_\mathrm{RPE}=100\,\mathrm{marcsec}$.

	\item $\mathbf{W}_{u_{fsm}}$ bounds the maximum available FSM input commands:
	\begin{equation}
		\mathbf{W}_{u_{fsm}} = \frac{1}{\bar{u}_{fsm}}\textbf{I}_2, \quad \text{with}\, \bar{u}_{fsm} = 5.3\,\mathrm{mrad} 
	\end{equation}
\end{itemize}

The robust 4-th order LOS observer is synthesized in an $\mathcal{H}_\infty$ framework using the non-smooth optimization algorithms \cite{apk2014,Apkarian2015} available in the MATLAB
routine \textit{systune}. This approach allows fixed structure low order controllers to be designed by imposing multi-objective optimization criteria and by coping with all parametric uncertainties in the model. 

The $\mathcal{H}_\infty$ optimization problem to find the optimal observer $\hat{\mathbf{K}}_{\mathrm{FSM}}(\mathrm s)$ is formulated as it follows:
\begin{align}
	\begin{split}
		&\hat{\mathbf{K}}_{\mathrm{FSM}}(\mathrm s) =  \underset{\mathbf K_{\mathrm{FSM}}(\mathrm s)}{\mathrm{argmin}} \underset{\bm{\Delta_\Omega},\bm{\Delta_\tau},\bm{\Delta_{A_\bullet}}}{\mathrm{max}}  \left\{
		\begin{array}{c}
			\gamma_1 = \left|\left| \mathcal{F}_l(\mathcal{P}(\mathrm s,\bm{\Delta_\Omega},\bm{\Delta_\tau},\bm{\Delta_{A_\bullet}}),{\mathbf{K}}_{\mathrm{FSM}})_{
				\left[
					\tilde{\mathbf{w}}_{rws}^\mathrm{T}\,
					\tilde{\mathbf{w}}_{sa}^\mathrm{T}
				\right]^\mathrm{T}\rightarrow
				{\mathbf{e}}_\mathrm{APE}}\right|\right|_\infty\\
			\gamma_2 = \left|\left| \mathcal{F}_l(\mathcal{P}(\mathrm s,\bm{\Delta_\Omega},\bm{\Delta_\tau},\bm{\Delta_{A_\bullet}}),{\mathbf{K}}_{\mathrm{FSM}})_{
				\left[
					\tilde{\mathbf{w}}_{rws}^\mathrm{T}\,
					\tilde{\mathbf{w}}_{sa}^\mathrm{T}
				\right]^\mathrm{T}\rightarrow
				{\mathbf{e}}_\mathrm{RPE}}\right|\right|_\infty	
		\end{array} \right.\\
		& \textbf{such that:} \quad \gamma_3 = \underset{\bm{\Delta_\Omega},\bm{\Delta_\tau},\bm{\Delta_{A_\bullet}}}{\mathrm{max}}\left|\left| \mathcal{F}_l(\mathcal{P}(\mathrm s,\bm{\Delta_\Omega},\bm{\Delta_\tau},\bm{\Delta_{A_\bullet}}),{\mathbf{K}}_{\mathrm{FSM}})_{
			\left[
				\tilde{\mathbf{w}}_{rws}^\mathrm{T}\,
				\tilde{\mathbf{w}}_{sa}^\mathrm{T}
			\right]^\mathrm{T}\rightarrow
			{\mathbf{e}}_u^\mathrm{fsm}}\right|\right|_\infty  < 1
	\end{split}
	\label{eq:fsm_synt}
\end{align}

\begin{figure}[!h]
	\centering
	\includegraphics[width=\columnwidth]{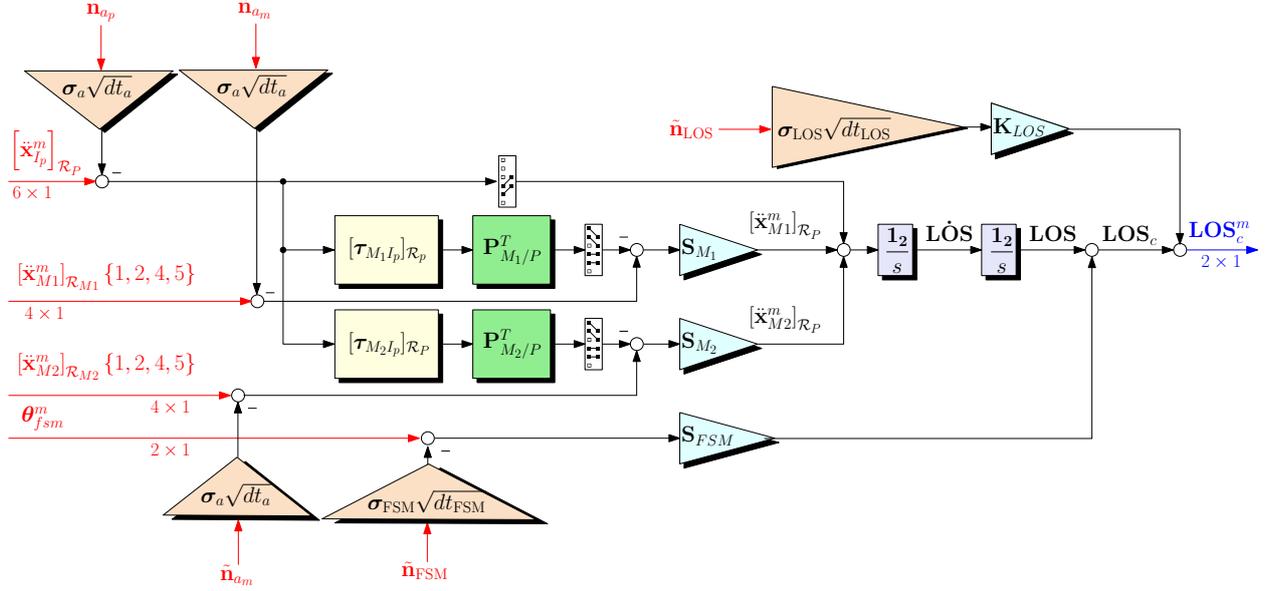}
	\caption{Kalman model for LOS estimation}
	\label{kalman model}
\end{figure}

Due to the gradient-based optimization nature of the non-smooth algorithm, it is necessary to find a good initial guess in order to get satisfactory results. This is why before running the $\mathcal{H}_\infty$ synthesis, a linear Kalman filter is synthesized as first guess of $\mathbf{K}_{FSM}(\mathrm s)$. The simplified model used for the Kalman filter design is shown in Fig. \ref{kalman model}: the controlled LOS $\textbf{LOS}_c$ is modelled by reconstructing the optical path from the accelerometers, the FSM strain gauges and the kinematic models between the sensor locations in the structure and is measured by the CCD camera. 

The primary measurements vector (input to the Kalman model) is composed by the inertial measurements from the accelerometers placed on the various elements of the payload, as well as the FSM $x$ and $y$ deflections measurements:

\begin{itemize}
	\item $\left[\ddot{\mathbf{x}}_{I_p}^m\right]_{\mathcal{R}_{P}}$ is the acceleration twist measurement at the payload isolator reference point projected in the payload frame $\mathcal R_p$,
	\item $\left[\ddot{\mathbf{x}}^m_{M1}\right]_{\mathcal{R}_{M1}}$ is the acceleration twist measurement at the point of the mirror $M_1$ projected in the mirror $M_1$ local frame $\mathcal R_{M_1}$,
	\item $\left[\ddot{\mathbf{x}}^m_{M2}\right]_{\mathcal{R}_{M2}}$ is the acceleration twist at the point of the mirror $M_2$ projected in the mirror $M_2$ local frame $\mathcal R_{M2}$,
	\item $\bm\theta_{fsm}^{m}$ is the angular position vector measurement of the FSM around the $\mathbf{x}$ and $\mathbf{y}$ axes in the payload frame $\mathcal R_p$.
\end{itemize}

The secondary measurements vector (output of the Kalman filter) is only composed of the LOS reconstruction done via image processing algorithm:

\begin{itemize}
	\item $\mathbf{LOS}_c = \mathbf{LOS}+\mathbf{S}_{FSM}\bm\theta_{fsm}$ is the controlled LOS.
\end{itemize}

The Kalman model contains the following kinematic parameters:

\begin{itemize}
	\item $[\bm\tau_{M_iI_p}]_{\mathcal R_p}$ is the kinematic model between the point $I_p$ and $M_i$ (expressed in the frame $\mathcal R_p$),
	\item $\mathbf P_{M_i/p}$ is the DCM mapping a vector expressed in the frame $\mathcal R_{M_i}$ into a vector expressed in the frame $\mathcal R_p$,
	\item $\mathbf S_{M_i}$ is the $2\times 6$ sensibility matrix of the LOS in the payload body frame $\mathcal R_p$ to the local deflections of the payload at the point $M_i$.
\end{itemize}

The normalized noise inputs $\tilde{\mathbf{n}}_\bullet$ are Gaussian centered white noises with unit power spectral density (PSD). 

This Kalman model does not take into account the low-pass behavior of the camera measurements in order to limit the order of the resulting estimator. For this reason the gain $\mathbf{K}_{LOS}=10^5$ is introduced to degrade the secondary measurement.
 The resulting filter, denoted $\mathbf{K}_{est}(\mathrm{s})\in \mathbb{R}^{2\times18}$, is a 4th order linear Kalman filter.

\subsection{Hybrid control of LOS with FSM and PMAs}
\label{sec:FEM+PMA}

The FSM control loop presented in section \ref{sec:fsm_control} manages to achieve a broadband rejection of microvibrations. A further reduction of the LOS error can be obtained if a set of six PMAs, connected to the payload isolator as shown in section \ref{sec:pma}, is used in an active isolation feedback loop on the payload acceleration $\ddot{\mathbf{x}}_{I_p}^m$. The control architecture is the one proposed in Fig. \ref{fig:synthesis_PMA}, where the generalized plant is now $\mathcal{Q}(s,\bm{\Delta_\Omega},\bm{\Delta_\tau},\bm{\Delta_{A_\bullet}})$ and the $6\times 6$ controller to be optimized is $\mathbf{K}_\mathrm{PMA}$.

\begin{figure}[!h]
	\centering
	\includegraphics[width=.8\columnwidth]{synthesis_PMA}
	\caption{Control architecture for PMA robust control synthesis}
	\label{fig:synthesis_PMA} 
\end{figure}

In this case, a finer RPE requirement than in FSM synthesis is imposed in order to exploit the high-pass filter behavior of the PMAs. 
We use then the filter:

	\begin{equation}
		\mathbf{W}_\mathrm{RPE}^f(\mathrm s) = \epsilon_{\mathrm{RPE}^f}^{-1} \frac{t_\Delta \mathrm s\left(t_\Delta \mathrm s + \sqrt{12}\right)}{\left(t_\Delta \mathrm s\right)^2+6\left(t_\Delta \mathrm s\right)+ 12}\textbf{I}_2
	\end{equation}
where $\epsilon_{\mathrm{RPE}^f} = 40\,\mathrm{marcsec}$  and $t_\Delta = 20\,\mathrm{ms}$. 
Moreover, the output filter $\mathbf{W}_{u_{pma}}$ overbounds the maximum PMA input force:
\begin{equation}
	\mathbf{W}_{u_{pma}} = \frac{1}{\bar u_{pma}}\textbf{I}_6
\end{equation}
with $\bar u_{pma} = 31.6\,\mathrm{N}$.

The mixed $\mathcal{H}_\infty/\mathcal{H}_2$ optimization problem to find the optimal controller $\hat{\mathbf{K}}_{\mathrm{PMA}}$ is formulated as it follows:
\begin{align}
	\begin{split}
		&\hat{\mathbf{K}}_{\mathrm{PMA}}(\mathrm s) =  \underset{\mathbf K_{\mathrm{PMA}}(\mathrm s)}{\mathrm{argmin}} \underset{\bm{\Delta_\Omega},\bm{\Delta_\tau},\bm{\Delta_{A_\bullet}}}{\mathrm{max}}  \left\{
		\begin{array}{c}
			\gamma_1 = \left|\left| \mathcal{F}_l(\mathcal{Q}(\mathrm s,\bm{\Delta_\Omega},\bm{\Delta_\tau},\bm{\Delta_{A_\bullet}}),{\mathbf{K}}_{\mathrm{FSM}})_{
				\left[
				\tilde{\mathbf{w}}_{rws}^\mathrm{T}\,
				\tilde{\mathbf{w}}_{sa}^\mathrm{T}
				\right]^\mathrm{T}\rightarrow
				{\mathbf{e}}_\mathrm{APE}}\right|\right|_\infty\\
			\gamma_2 = \left|\left| \mathcal{F}_l(\mathcal{Q}(\mathrm s,\bm{\Delta_\Omega},\bm{\Delta_\tau},\bm{\Delta_{A_\bullet}}),{\mathbf{K}}_{\mathrm{PMA}})_{
				\left[
				\tilde{\mathbf{w}}_{rws}^\mathrm{T}\,
				\tilde{\mathbf{w}}_{sa}^\mathrm{T}
				\right]^\mathrm{T}\rightarrow
				{\mathbf{e}}_\mathrm{RPE}^{f}}\right|\right|_\infty	\\
			\gamma_3 = \left|\left| \mathcal{F}_l(\mathcal{Q}(\mathrm s,\bm{\Delta_\Omega},\bm{\Delta_\tau},\bm{\Delta_{A_\bullet}}),{\mathbf{K}}_{\mathrm{PMA}})_{\tilde{\mathbf{n}}_{a_p}\rightarrow
				{\mathbf{e}}_\mathrm{APE}}\right|\right|_2
		\end{array} \right.\\
		& \textbf{such that:} \quad \gamma_4 = \underset{\bm{\Delta_\Omega},\bm{\Delta_\tau},\bm{\Delta_{A_\bullet}}}{\mathrm{max}}\left|\left| \mathcal{F}_l(\mathcal{Q}(\mathrm s,\bm{\Delta_\Omega},\bm{\Delta_\tau},\bm{\Delta_{A_\bullet}}),{\mathbf{K}}_{\mathrm{PMA}})_{
			\left[
			\tilde{\mathbf{w}}_{rws}^\mathrm{T}\,
			\tilde{\mathbf{w}}_{sa}^\mathrm{T}
			\right]^\mathrm{T}\rightarrow
			{\mathbf{e}}_u^\mathrm{pma}}\right|\right|_\infty  < 1
	\end{split}
	\label{eq:pma_synt}
\end{align}

Note that the $\mathcal{H}_2$-norm objective $\gamma_3$ is considered in order to limit the amplification of the accelerometers noise by the minimization of the variance between measurement noise and pointing performance.

\section{Results and discussion}
\label{results_discussion}

In this section, the results obtained with the control architectures outlined in section \ref{sec:fsm_control} and \ref{sec:FEM+PMA} are analyzed. 
Table \ref{tab:FSM_PMA_synthesis} resumes the achieved optimization performance indexes.
One can notice that for the FSM robust control synthesis (optimization problem \eqref{eq:fsm_synt}), the limits of system performance are reached since all indexes reach the unity value). The hard constraint on control authority $\gamma_3$ is satisfied by leaving a very small margin for further improvement of the two pointing performance indexes, $\gamma_1$ and $\gamma_2$.  Note that the RPE performance index $\gamma_2$ is slightly bigger than unity, fact that is not considered critical for this design. Moreover note that these indexes corresponds to the worst-case achievable performance by taking into account all possible uncertainties and considering the biggest $\mathcal{H}_\infty$-norm on all parametric configurations. The saturation of the three performance indexes means that a further reduction of the LOS jitter cannot be demanded since the limit of performance is already been reached. This limit is in fact imposed both by the physical characteristics (maximum accepted FSM input signal, noise) of the set of sensor/actuators chosen for this control architecture and the need of performance robustness against uncertain/variable parameters.

In order to further improve the jitter rejection, a second stage of micro-vibration active control is added to the already synthesized FSM closed-loop. A set of 6 PMA is then introduced in order to reject the transmitted disturbances to the payload base and improve the performance reached by the FSM. For the synthesis of the PMA controller the optimization problem \eqref{eq:pma_synt} is solved. Note that this time a reduction of 60 marcsec is asked in terms of RPE performance with respect to the FSM closed-loop synthesis. As shown in Table \ref{tab:FSM_PMA_synthesis} the more restrictive requirement on RPE performance ($\gamma_2$) is met. However only $\approx 12\%$ of the available control signal ($\gamma_4$) is sufficient to guarantee the requested pointing performance ($\gamma_1$ and $\gamma_2$) by coping with a limitation of the accelerometer noise propagation to the LOS ($\gamma_3$). This means that a set of PMA with lower maximum available input force could be sufficient to meet the same level of pointing performance.
Moreover, one can notice that the APE requirement is largely met ($\gamma_1 = 0.5768$) since the FSM control stage already takes care of it. The biggest limitation in this design is then due to the noise introduced by the chosen set of accelerometers ($\gamma_3$).

\begin{table}[!h]
	\centering
	\begin{tabular}{ccc|cccc}
		\hline
		\multicolumn{3}{c}{\textbf{FSM Synthesis}} & \multicolumn{4}{c}{\textbf{PMA Synthesis}} \\ \hline
		$\gamma_1$ & $\gamma_2$ & $\gamma_3$ & $\gamma_1$ & $\gamma_2$ & $\gamma_3$ & $\gamma_4$ \\ \hline
		0.9857 & 1.0026 & 0.9877 & 0.5768 & 0.9835 & 0.9835 & 0.1185 \\
		\hline
	\end{tabular}
	\caption{\textcolor{black}{Worst-case performance of FSM and PMA robust control design}}
	\label{tab:FSM_PMA_synthesis}
\end{table}

\begin{figure}[!h]
	\centering
	\begin{subfigure}[b]{0.49\columnwidth}
		\centering
		\includegraphics[width=\columnwidth]{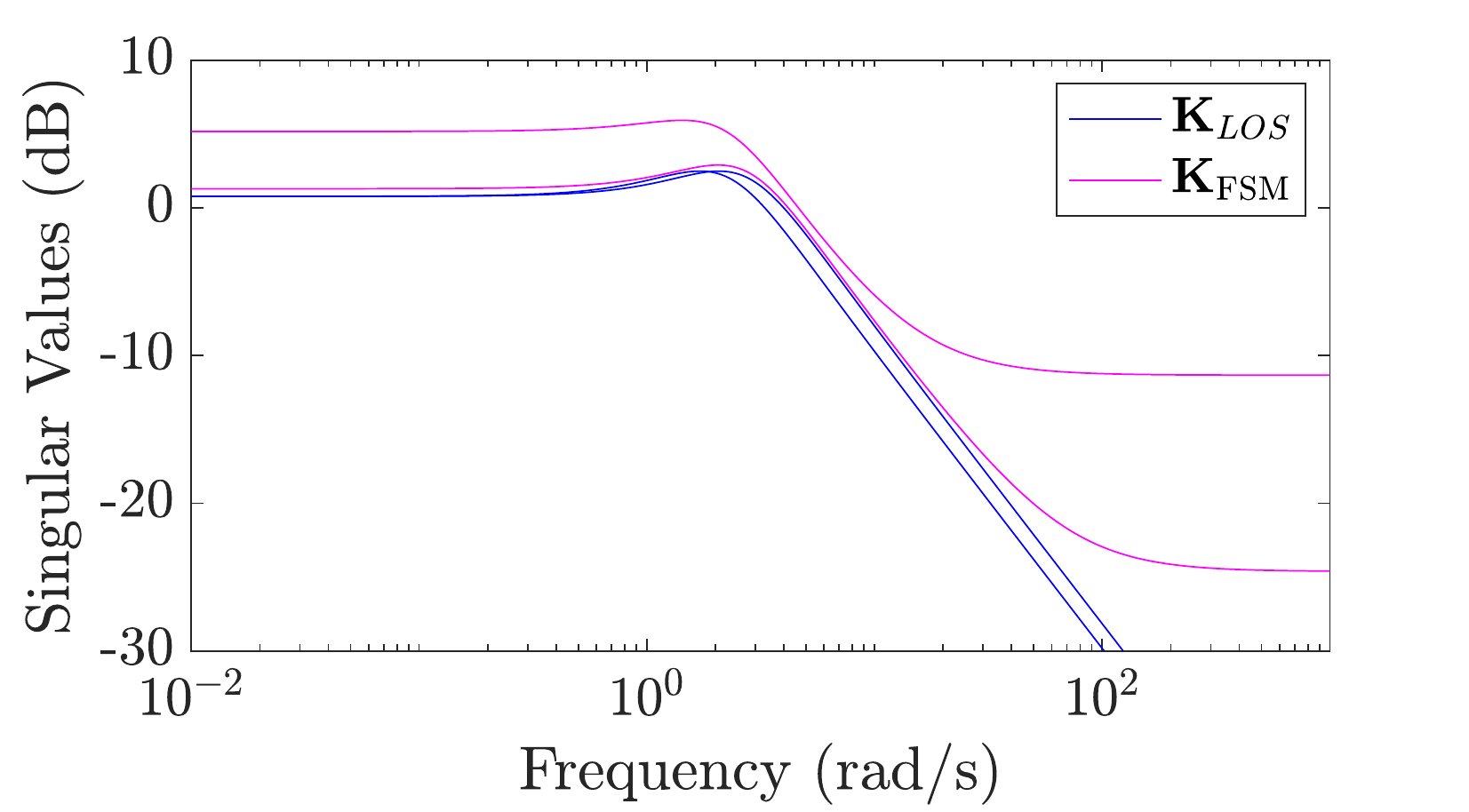}
		\caption{}
		\label{fig:k_LOS_k_FSM}
	\end{subfigure}
	\begin{subfigure}[b]{0.49\columnwidth}
		\centering
		\includegraphics[width=\columnwidth]{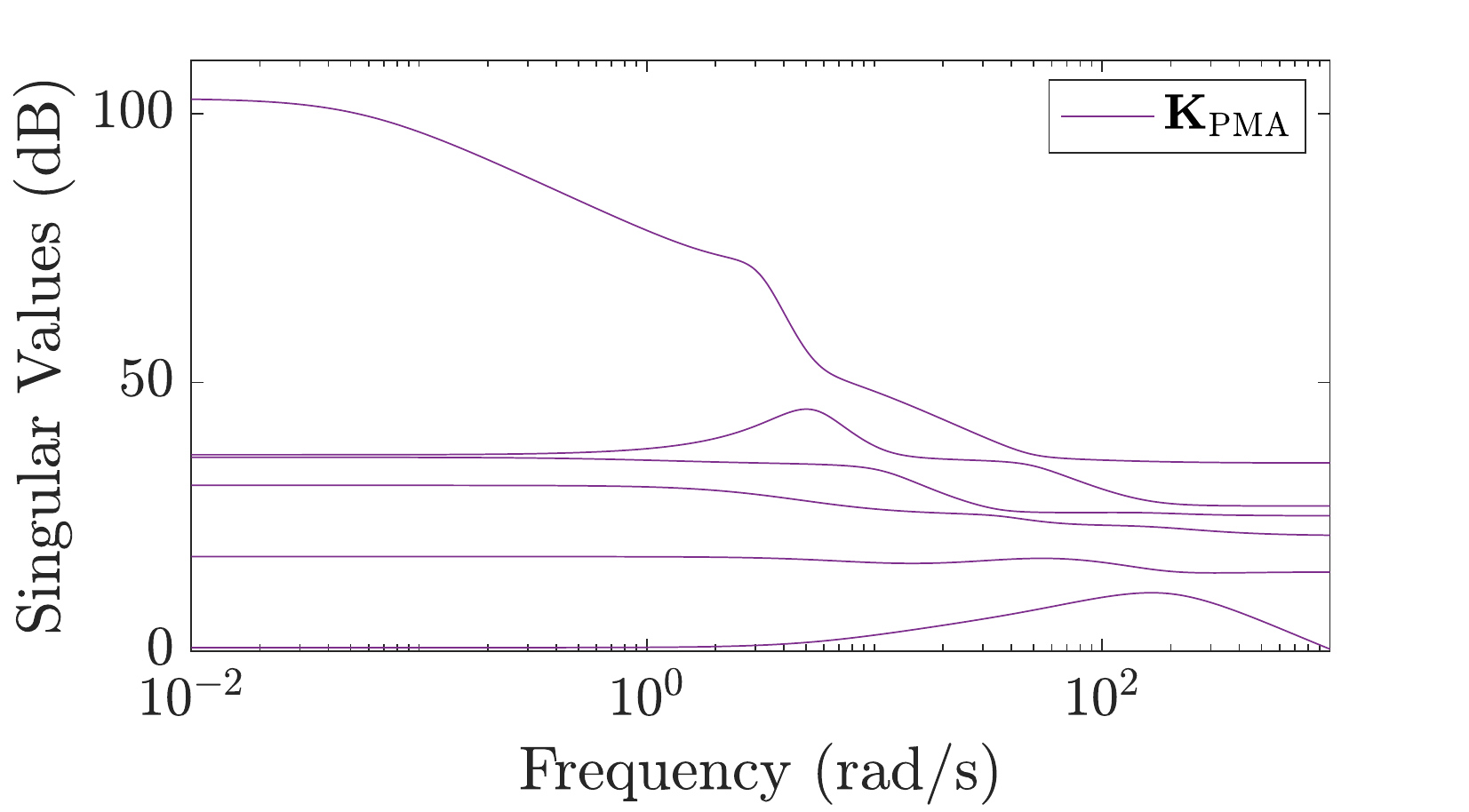}
		\caption{}
		\label{fig:k_PMA}
	\end{subfigure}
	\caption{Fine LOS controllers: (a) Kalman filter first guess $\mathbf{K}_{LOS}$ and optimal FSM controller $\mathbf{K}_\mathrm{FSM}$; (b) optimal PMA controller $\mathbf{K}_\mathrm{PMA}$}
	\label{fig:controllers}
\end{figure}

Figure \ref{fig:controllers} shows the singular values of the two optimal controllers. Figure \ref{fig:k_LOS_k_FSM} compares the final $\hat{\mathbf{K}}_\mathrm{FSM}(\mathrm s)$ with the initial Kalman Filter first guess $\mathbf{K}_{est}(\mathrm s)$. The FSM controller is of course a 4th order system as its corresponding Kalman Filter.
For the synthesis of the PMA controller $\mathbf{K}_\mathrm{PMA}(\mathrm s)$, several fixed order system have been tested and finally a 4-th order structure demonstrated to be sufficient to obtain the required level of performance without no remarkable performance improvement for higher order. A random start option has been used with the MATLAB routine \textit{systune} to get an optimal first guess.

\begin{figure}[h!]
	\centering
	\includegraphics[width=\columnwidth]{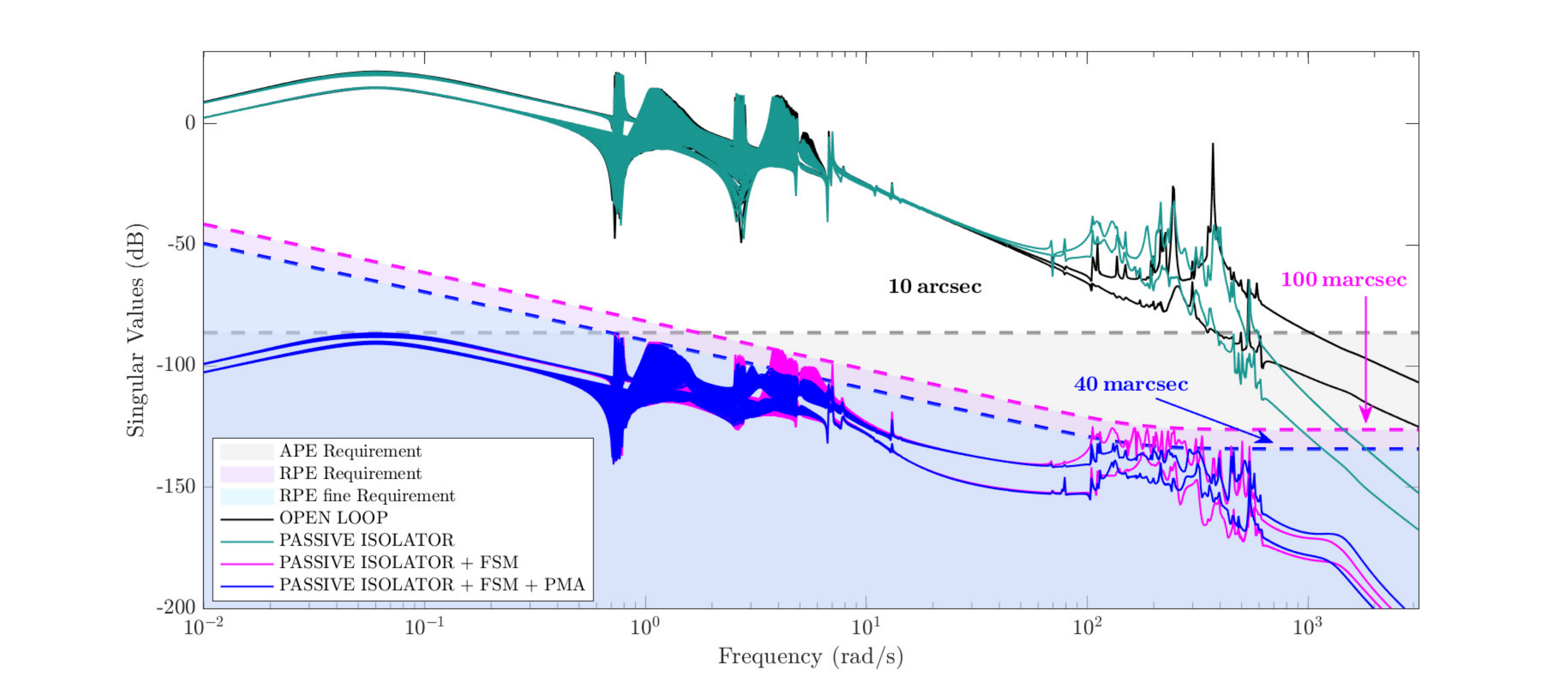}
	\caption{Singular Values of the transfer function $\left[\begin{array}{cc}
			\tilde{\mathbf{w}}_{rws}^\mathrm{T} &
			\tilde{\mathbf{w}}_{sa}^\mathrm{T}
		\end{array}\right]^\mathrm{T}\rightarrow \textbf{LOS}_c$ \textcolor{black}{with robust control design}. Comparison of pointing performance achieved with just a payload passive isolator (green line), with a payload isolator and an FSM (magenta line) and with a payload isolator, an FSM and a set of six PMAs (blue line)}
	\label{dist2LOS}
\end{figure}

The optimal design of the two-stages active micro-vibration control can be also visualized in the frequency domain, where the worst-case disturbance rejection can be analyzed as a function of the frequency.
For this purpose, Fig. \ref{dist2LOS} shows the singular values of the transfer function from all normalized input disturbance signals (both RW and SADM perturbation) to the controlled LOS.
Four sets of curves are depicted in order to compare the pointing performance reached by increasing the number of adopted micro-vibrations control stages: the black lines show the transmissibility of the microvibration to the LOS in open-loop when neither passive (no isolation at payload level) nor active solutions are used; green lines depicts the same transfer functions when a passive isolator platform is introduced at the base of the payload and no active LOS control is used; magenta lines corresponds to the configuration combining the passive payload isolator together with the FSM active control stage; blue lines finally relate to the final configuration with passive isolator and double-stage LOS active control with FSM and PMAs on the isolator platform. 
Note that the line clouds of the same color correspond to different samples of the same uncertain closed-loop system.
In the same Fig. \ref{dist2LOS} the APE and RPE specifications are illustrated as well. They all correspond to  the inverse of the output filters $\mathbf{W}_\mathrm{APE}$, $\mathbf{W}_\mathrm{RPE}$ and $\mathbf{W}_\mathrm{RPE}^f$.

When a passive isolator is connected at the base of the payload, an important reduction of the propagation of the disturbance through the optical path is achieved for frequencies bigger than $\approx 500\,\mathrm{rad/s}$ (green line). As already seen in Fig. \ref{fig:passiveVsOL}, this strategy allows for a reduction of micro-vibration propagation at very high frequency while introducing some extra modes at lower frequencies due to the isolator stiffness. Only an active control strategy can then reduce the impact of these modes.
The use of an FSM drastically improves the pointing performance by dropping the APE below $10\,\,\mathrm{arcsec}$ and the RPE below $100\,\mathrm{marcsec}$ for frequencies above $100\,\mathrm{rad/s}$. A further improvement of the RPE is achieved with the use of a set of PMAs connected to the payload isolator (blue line) with a gain of almost $8\,\mathrm{dB}$ along all frequencies and guarantees a LOS error below $40\,\mathrm{marcsec}$ for frequencies above $200\,\mathrm{rad/s}$.

\begin{figure}[h!]
	\centering
	\includegraphics[width=\columnwidth]{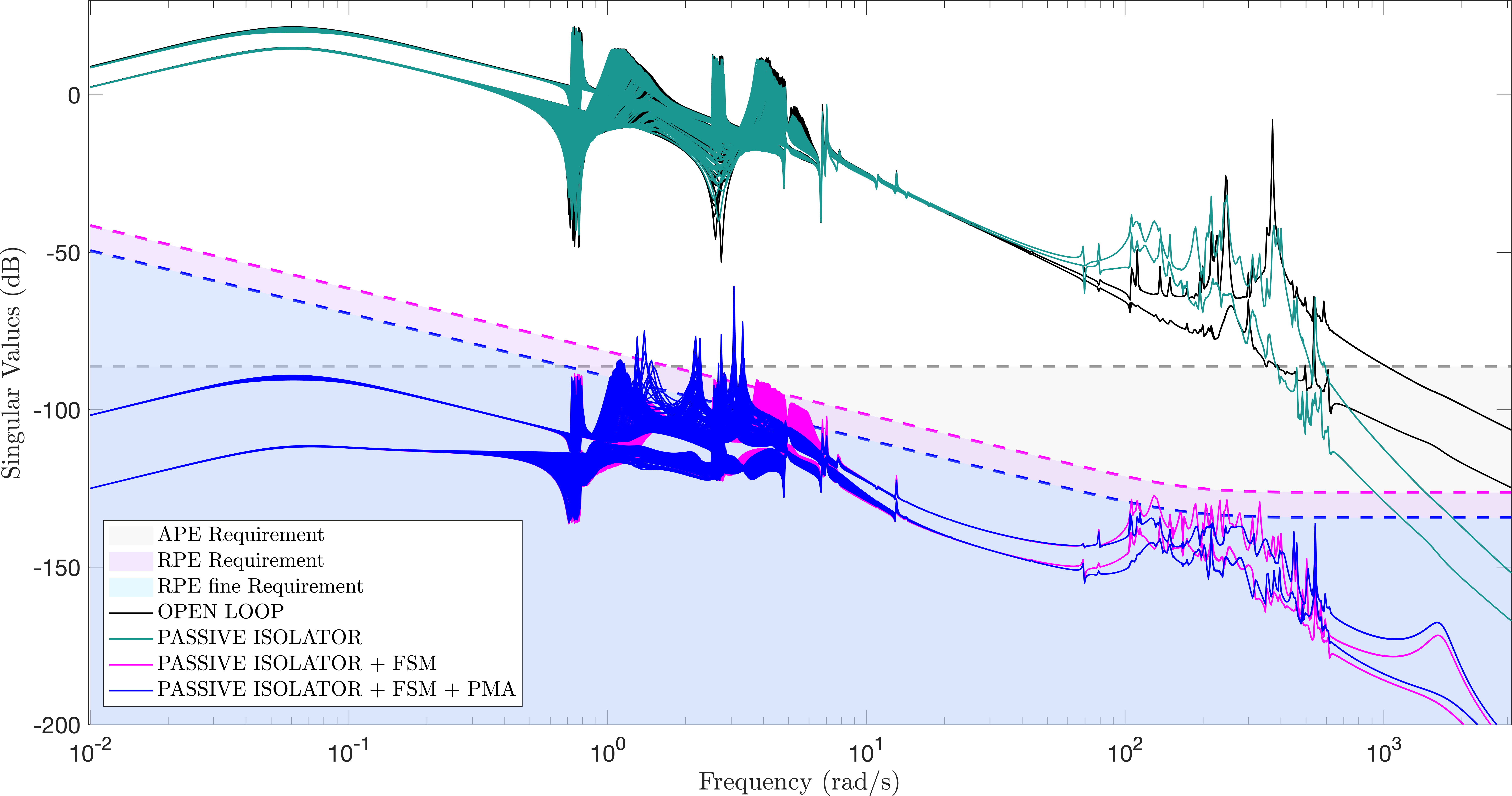}
	\caption{Singular Values of the transfer function $\left[\begin{array}{cc}
			\tilde{\mathbf{w}}_{rws}^\mathrm{T} &
			\tilde{\mathbf{w}}_{sa}^\mathrm{T}
		\end{array}\right]^\mathrm{T}\rightarrow \textbf{LOS}_c$ with nominal control design. Comparison of pointing performance achieved with just a payload passive isolator (green line), with a payload isolator and an FSM (magenta line) and with a payload isolator, an FSM and a set of six PMAs (blue line)}
	\label{dist2LOS_Nominal}
\end{figure}

It is to be stressed that all these results are robustly guaranteed for any solar array angular configuration, RW speed and modeled uncertainty on the first two SA flexible modes when both RW and SADM are kept activated during the imaging phase. 
\textcolor{black}{In order to highlight the interest of considering parametric models from the early preliminary design phases, the FSM and PMA controllers were designed with the same approach presented in Sections
\ref{sec:fsm_control} and \ref{sec:FEM+PMA} but with the nominal model without any uncertainties or variable parameters. Despite the nominal performances  indexes summarized in Table \ref{tab:FSM_PMA_synthesis_nominal} are very good, the validation of this nominal controller on the uncertain system failed as shown in Fig. \ref{dist2LOS_Nominal}: some parametric configurations lead to high magnitude peaks outside the requirement templates.}

\begin{table}[!h]
	\centering
	\begin{tabular}{ccc|cccc}
		\hline
		\multicolumn{3}{c}{\textbf{FSM Synthesis}} & \multicolumn{4}{c}{\textbf{PMA Synthesis}} \\ \hline
		$\gamma_1$ & $\gamma_2$ & $\gamma_3$ & $\gamma_1$ & $\gamma_2$ & $\gamma_3$ & $\gamma_4$ \\ \hline
		0.7115 & 0.7160 & 0.9804 & 0.7115 & 0.7115 & 0.6150 & 0.4266 \\
		\hline
	\end{tabular}
	\caption{Nominal performance of FSM and PMA control design}
	\label{tab:FSM_PMA_synthesis_nominal}
\end{table}

According to the achieved results, some lesson learned can be summed up:
\begin{itemize}
    \item The multi-body TITOP approach allowed us to easily build an industrial benchmark by connection of some elementary blocks. From a preliminary design point of view, this assembling strategy is convenient to analyze the transmissibility of vibrations at any point of a complex system and possibly propose a different system layout by easily displacing/adding/removing the constitutive sub-structures;
    \item The TITOP modeling approach facilitates the choice of the set of sensors/actuators in a preliminary design phase by easily checking their efficiency at different points of the structure in the same way as described in the previous point;
    \item The analytical dependency of the TITOP models on their physical mechanical parameters with the LFT formalism allows the user to directly synthesize robust control laws by taking into account all the possible worst-case scenarios. This point facilitates the successive analysis of the system, that can be formally validated with analytical guarantees of system stability and performance \cite{roos2011} without involving time-expensive Monte Carlo campaigns;
    \item The proposed double-stage active micro-vibration rejection strategy allowed us to achieve very fine pointing performance. It combined a direct correction of the LOS with an FSM located in front of the sensitive instrument and a set of PMAs acting on a passive isolator, that mitigate the propagation of the microvibrations produced by the RWs and the SADM to the payload base through the most flexible elements of the spacecraft (solar panels, large mirrors of the telescope).
\end{itemize}

\section{Conclusion}
\label{conclusion}
An analytical methodology to model all flexible elements and mechanisms of a scientific satellite and its optical payload in a multi-body framework was presented. In particular the Two-Input Two-Output Ports approach was used to propose novel models for a reaction wheel assembly including its imbalances and two kinds of actuators to control the line-of-sight: an FSM and a set of PMAs. This approach allowed the authors to assemble a complex industrial spacecraft where detailed finite element models can be easily included as well. Since in this framework an uncertain Linear Parametric-Varying system can be directly derived by including all possible configurations and uncertainties of the plant, two novel robust active control strategies have been proposed to mitigate the propagation of the microvibrations to the LOS error. A first one consists in synthesizing an observer of the LOS error by blending the low-frequency measurements of the LOS directly provided by a CCD camera and the accelerations measured in correspondence of the most flexible optical elements (mirrors $M_1$ and $M_2$ of a space telescope) together with the accelerations measured on a passive isolator placed at the base of the payload.  An FSM then uses this information to mitigate the pointing error. In order to obtain even tighter micro-vibration attenuation, a second stage of active control was proposed as well. This strategy consists in measuring the accelerations of the payload isolator and actuating six PMAs attached to the same isolator. Thanks to this double-stage active control strategy, the propagation of the micro-vibrations induced by the RWs and SADMs is finely reduced on a very large frequency band. In particular, a reduction of the pointing error to 10 arcsec is guaranteed at low frequency ($\approx 1$ rad/s) with a progressive reduction of the jitter until 40 marcsec for higher frequencies where micro-vibration sources act.

This application finally allowed the authors to demonstrate the interest of the proposed modeling approach, that is able to finely capture the dynamics of a complex industrial benchmark by including all possible uncertainties in a unique LFT model. This modular framework, which permits to easily build and design a multi-body flexible structure, was in fact conceived in order to perfectly fit with the modern robust control theory. In this way the authors demonstrated how to push the control design to the limits of achievable performance, which is fundamental in the preliminary design phases of systems with very challenging pointing requirements.

\section*{Funding Sources}

This research was founded by the ESA Open Invitations to Tender "Line Of Sight Stabilisation Techniques (LOSST)" performed together with Thales Alenia Space, Cannes, France (Contract NO. 1520095474 / 02).


\appendix
\section{Spacecraft Data}
See Table \ref{tab:data}.
\begin{table}[!h]
	\renewcommand{\arraystretch}{1.3}
	%\caption{Values of the spacecraft parameters.}
	\label{tab:spacecraft_data_sadm}
	\centering
	\begin{footnotesize}
		\resizebox{\textwidth}{!}{\begin{tabular}{cllr}
				\hline
				\textbf{System} & \textbf{Parameter} & \textbf{Description} & \textbf{Nominal Value } \\ \hline
				& $m^{\mathcal{A}_\bullet}$ & Mass & $59\,\mathrm{kg}$\\
				&  $\left[\begin{array}{ccc}
					J_{xx}^{\mathcal{A}_\bullet} & J_{xy}^{\mathcal{A}_\bullet} & J_{xz}^{\mathcal{A}_\bullet} \\
					& I_{yy}^{\mathcal{A}_\bullet} & I_{yz}^{\mathcal{A}_\bullet} \\
					sym & & I_{zz}^{\mathcal{A}_\bullet}
				\end{array}\right]$ & Inertia in $\mathcal{R}_{a_\bullet}$ frame &  $\left[\begin{array}{ccc}
					443.49& 0.36 & -0.01 \\
					& 26.14 & 0.09 \\
					sym & & 469.62
				\end{array}\right]\mathrm{kg\, m^2}$\\
				& $\mathbf{r}_{o}^{\mathcal{A}_\bullet}$ & $\mathcal{A}_\bullet$ CG in $\mathcal{R}_{a_\bullet}$ frame &  $\left[0.007\,\,-5.517\,\,0\right]\,\mathrm{m}$\\
				& $\left[\omega_1^{\mathcal{A}_\bullet}\, \omega_2^{\mathcal{A}_\bullet}\,\dots\right]$ & Flexible modes' frequencies & $\left[0.743 \pm 5 \%\,\,2.65\pm 5 \% \dots \right]\,\mathrm{rad/s}$ \\
				& $\zeta_1^{\mathcal{A}_\bullet}, \zeta_2^{\mathcal{A}_\bullet}, \dots$ & Flexible modes' damping & $0.001$\\
				\multirow{-10}{*}{\shortstack{Solar \\ Array $\mathcal{A}_\bullet$}} & $\mathbf{L}_P^{\mathcal{A}_\bullet}$ & Modal participation factors & $\left[\begin{array}{cccccc}
					-0.0001 & 0.0051 & 6.5226 & -47.0158 & -0.0546 &  -0.0008\\
					6.8249 & 0.0090 & 0.0003 & -0.0012 & 0.0279 &  47.5976 \\
					\vdots & \vdots & \vdots & \vdots & \vdots & \vdots \\
				\end{array}\right]$ \\ \hline
			& $m^{\mathcal{IA}}$ & Mass & $1.50\,\mathrm{kg}$\\
			&  $\left[\begin{array}{ccc}
				J_{xx}^{\mathcal{IA}} & J_{xy}^{\mathcal{IA}} & J_{xz}^{\mathcal{IA}} \\
				& I_{yy}^{\mathcal{IA}} & I_{yz}^{\mathcal{IA}} \\
				sym & & I_{zz}^{\mathcal{IA}}
			\end{array}\right]$ & Inertia in $\mathcal{R}_{ia}$ frame &  $\left[\begin{array}{ccc}
				0.41& -0.06 & 0 \\
				& 0.58 & 0 \\
				sym & & 0.99
			\end{array}\right]\mathrm{kg\, m^2}$\\
			\multirow{-3}{*}{\shortstack{Payload \\ Isolator \\$\mathcal{IA}$}}& $\mathbf{r}_{o}^{\mathcal{IA}}$ & $\mathcal{IA}$ CG in $\mathcal{R}_{ia}$ frame &  $\left[-0.046\,\,0.039\,\,0.023\right]\,\mathrm{m}$\\
				& $\mathbf{K}^\mathcal{IA}$ & Stiffness  & $\mathrm{blkdiag}(5.32\,\mathrm{N/m}\,\textbf{I}_3,1.46\,\mathrm{Nm/rad},2.08\,\mathrm{Nm/rad},3.53\,\mathrm{Nm/rad})\cdot 10^6$\\
				& $\mathbf{D}^\mathcal{IA}$ & Damping  & $\mathrm{blkdiag}(1.1781\,\mathrm{Ns/m}\,\textbf{I}_3,\,1.1781\,\mathrm{Ns/rad}\,\textbf{I}_3)\cdot 10^3$\\ \hline 
				& $m^{\mathcal{RW}}$ & Mass & $1\,\mathrm{kg}$\\
				& $J_w^{\mathcal{RW}}$ & Radial inertia in $\mathcal{R}_{w}$ frame& $0.047\,\mathrm{kg\,m^2}$\\
					\multirow{-2}{*}{\shortstack{Reaction \\ Wheel \\$\mathcal{RW}$}}& $J_r^{\mathcal{RW}}$ & Axial inertia in $\mathcal{R}_{w}$ frame& $0.096\,\mathrm{kg\,m^2}$\\
				& ${\Omega}^\mathcal{RW}$ & Speed range  & $\left[-1.0472,\,1.0472\right]\cdot 10^3\,\mathrm{rad/s}$\\ \hline 
				& $m^{\mathcal{FSM}}$ & Mass & $0.05\,\mathrm{kg}$\\
				& $J^{\mathcal{FSM}}$ & Inertia in $\mathcal{R}_{fsm}$ frame& $\mathrm{diag}(0.313,0.313,0.625)\cdot 10^{-4}\,\mathrm{kg\,m^2}$\\
				\multirow{-2}{*}{\shortstack{Fast Steering \\ Mirror \\$\mathcal{FSM}$}}& $\mathbf{k}^{\mathcal{FSM}}$ & Stiffness & $77.11\,\textbf{I}_2\,\mathrm{N\,m/rad}$\\
				& $\mathbf{d}^{\mathcal{FSM}}$ & Damping & $0.015,\textbf{I}_2\,\mathrm{N\,s/rad}$\\ \hline 
					& $m^{\mathcal{PMA}}$ & Moving mass & $2.5\,\mathrm{kg}$\\
					& $M^{\mathcal{PMA}}$ & Caging mass & $3.5\,\mathrm{kg}$\\
				& $J^{\mathcal{PMA}}$ & Inertia in $\mathcal{R}_{pma}$ frame& $\mathrm{diag}(7.40,\,7.40,\,3.15)\cdot 10^{-4}\,\mathrm{kg\,m^2}$\\
				\multirow{-2}{*}{\shortstack{Proof-Mass \\ Actuator \\$\mathcal{PMA}$}}& $k^{\mathcal{PMA}}$ & Stiffness & $70\,\mathrm{N/m}$\\
				& $d^{\mathcal{PMA}}$ & Damping & $10\,\mathrm{N\,s/m}$\\ \hline 
		\end{tabular}}
	\end{footnotesize}
	\caption{Spacecraft data. Note that not provided data are confidential.}
	\label{tab:data}
\end{table}

%% References


\bibliographystyle{model1-num-names}
\bibliography{sample.bib}






% \end{thebibliography}

\end{document}